\documentclass[pdflatex,sn-mathphys-ay]{sn-jnl}

\usepackage{graphicx}%
\usepackage{multirow}%
\usepackage{amsmath,amssymb,amsfonts}%
\usepackage{amsthm}%
\usepackage{mathrsfs}%
\usepackage[title]{appendix}%
\usepackage{xcolor}%
\usepackage{textcomp}%
\usepackage{manyfoot}%
\usepackage{booktabs}%
\usepackage{quantikz}
\usepackage{algorithm}%
\usepackage{algorithmicx}%
\usepackage{algpseudocode}%
\usepackage{listings}%

\theoremstyle{thmstyleone}%
\theoremstyle{thmstyletwo}%

\theoremstyle{thmstylethree}%

\begin{document}

\title[Article Title]{An Efficient Quantum Circuit for Flow Model Execution Using Quantum Neural Networks}

\author*[1]{\fnm{Rui} \sur{Che}}\email{rui.che@mdu.se}

\author[1]{\fnm{Ludvig} \sur{af Klinteberg}}\email{ludvig.af.klinteberg@mdu.se}

\affil*[1]{\orgdiv{Department of Business and Mathematics}, \orgname{Mälardalen University}, \orgaddress{\street{Universitetsplan 1}, \city{Västerås}, \postcode{721 23}, \country{Sweden}}}

\abstract{Flow models generate trajectories from an initial distribution to a target distribution by solving an ordinary differential equation defined by a velocity field. Flow matching learns this velocity field by modeling the transport dynamics between the two distributions. Wavefunction flow establishes a formal connection between flow models and quantum dynamics by introducing a continuity Hamiltonian, which drives the Schr\"odinger evolution of quantum states. In this paper, we investigate accurate and efficient quantum simulation of the wavefunction flow, thereby realizing the efficient implementation of flow models on quantum computers. We first leverage a quantum read-only memory (QROM)-based phase kickback framework for the wavefunction flow simulation, generating probability densities that closely match those produced by the corresponding conventional flow model. To address the high circuit-resource cost, we further incorporate a trained quantum neural network (QNN) into the phase kickback framework, replacing QROM for data encoding. Numerical experiments demonstrate that our proposed method implements flow models on quantum computers more efficiently, since it maintains the accuracy of wavefunction flow simulation compared with the QROM-based framework, and significantly reduces the circuit resources.}




\maketitle

\section{Introduction}\label{sec1}

Generative models \citep{bond2021deep} are a class of machine learning algorithms that learn unknown data distributions and generate new samples accordingly. The most widely used generative models include Generative Adversarial Networks (GANs) \citep{goodfellow2014generative}, Variational Autoencoders (VAEs) \citep{kingma2013auto}, Autoregressive models \citep{van2016pixel}, and flow models \citep{gat2024discrete, lipman2022flow}. Flow models aim at generating complex target distributions from simple source distributions by learning velocity fields. The learned velocity fields govern the transportation of distributions. Flow matching \citep{lipman2022flow} provides an efficient method of learning velocity fields by approximating trajectories between source and target distributions that have been predefined, which is an ordinary differential equation (ODE)-based model. 

Recently, the connection between flow models and quantum dynamics has been developed \citep{layden2025wavefunction}. By introducing a continuity Hamiltonian $\hat{H}$, the implementation of flow models on quantum computers can be regarded as the Schr\"odinger evolution under this continuity Hamiltonian. The resulting quantum dynamics is defined as wavefunction flow. Wavefunction flow can be realized through the simulation of the Hamiltonian evolution operator $e^{-it\hat{H}}$. By discretizing the evolution time and using the Suzuki-Trotter formula, it can be expressed as a product of exponentiation operators $e^{-i\Delta t \hat{H}_j}$ corresponding to all Hamiltonian terms $\hat{H}_j$ \citep{chen2022efficient, low2023complexity}. Here, $\Delta t$ denotes the discretized time step interval. Each operator can be regarded as a quantum oracle and is implemented sequentially in the quantum circuit \citep{kornell2023some}. 

Phase kickback \citep{cleve1998quantum} is a classical quantum technique for the implementation of the evolution operator $e^{-i\Delta t H_j}$. It encodes the eigenvalues of $\hat{H}_j$ into ancilla qubits, then leverages rotation gates to generate the desired phase $e^{-i\Delta t \hat{H}_j}$. Finally, the adjoint eigenvalue encoding oracle is employed to uncompute the ancilla qubits to the initial states. In the wavefunction flow framework, the continuity Hamiltonian $\hat{H}_j$ is constructed by the time-dependent potential term $V_t$ and the kinetic term $K$. These two terms can be transformed into diagonal operators $D_{V_t}$ and $D_K$ by selecting the appropriate basis \citep{layden2025wavefunction}. To implement such evolution operators $e^{-i\Delta t D_{V_t}}$ and $e^{-i\Delta t D_K}$ using phase kickback frameworks, one of the key steps is to encode all diagonal entries of $D_{V_t}$ and $D_K$ into the quantum register. 

Quantum read-only memory (QROM) \citep{babbush2018encoding,lee2021even} is a widely used quantum technique for encoding classical data in the quantum circuit, where each data item corresponds to an initial basis state. We utilize QROM-based phase kickback frameworks to implement the evolution operators $e^{-i\Delta t D_{V_t}}$ and $e^{-i\Delta t D_K}$ for the wavefunction flow simulation, and verify the correspondence between the flow model and the wavefunction flow simulation. However, the QROM circuit is constructed by a series of multi-controlled $X$ gates \citep{zindorf2025efficient}, and the decomposition of a multi-controlled $X$ with $n$ control qubits without ancilla qubits requires $\Theta(2^n)$ CNOT gates \citep{barenco1995elementary}. Here, $\Theta$ denotes ``of order exactly $2^n$'' \citep{knuth1976big}. Hence, QROM faces exponentially increasing computational complexity, which leads to a high circuit resource cost for the wavefunction flow simulation.

To solve this problem, we introduce a specialized quantum neural network (QNN) circuit to replace QROM in the phase kickback framework for the wavefunction flow simulation. The QNN is a type of parameterized quantum circuit (PQC)-based variational quantum algorithm \citep{abbas2021power,schuld2014quest}. It combines features of neural networks and quantum circuits. The main areas of QNN application include classification problems \citep{mahmud2024quantum}, generative learning \citep{tian2023recent}, optimization \citep{amosy2024iteration}, chemistry \citep{smaldone2025quantum}, and financial predictions \citep{paquet2022quantumleap}. Through appropriate and sufficient training, our QNN-based phase kickback framework can maintain accuracy compared to that of the QROM-based framework. In addition, with this replacement, the number of required CNOT gates can be reduced to $\Theta(n)$.  QNN has been applied to Hamiltonian simulation using the variational quantum simulator (VQS) \citep{li2017efficient,heya2023subspace}. The VQS trains the QNN to learn the target quantum state from the given initial quantum state. However, each such QNN is designed for a single initial quantum state, and can generate only the corresponding target quantum state. For a diagonal Hamiltonian operator in the wavefunction flow model, where each diagonal entry corresponds to a different initial state, VQS is unable to generate all the desired target states by training the QNN. Therefore, this approach is not capable of realizing the evolution generated by the entire Hamiltonian. In contrast, our QNN-based framework achieves the whole evolution efficiently.

In this paper, inspired by the theory of wavefunction flow \citep{layden2025wavefunction}, we investigate an efficient simulation framework for wavefunction flow, thereby realizing the implementation of flow models on quantum computers efficiently. Building on the QROM-based phase kickback framework, we develop a QNN circuit to replace the QROM for data encoding in the wavefunction flow simulation. The new approach significantly reduces the cost of circuit resources while maintaining accuracy compared with the QROM-based framework. The contributions of the paper are as follows.
\begin{itemize}
    \item We leverage a QROM-based phase kickback technique to simulate wavefunction flow, thereby verifying the wavefunction flow theory and demonstrating the feasibility of implementing flow models on quantum computers. 
    \item We propose a QNN circuit to replace QROM in the phase kickback framework for data encoding in the wavefunction flow simulation. Instead of employing multi-controlled X gates to encode the exact binary bits in the ancilla qubits through QROM, our QNN circuit learns the mapping from the states of the address qubits to the corresponding probabilities of the ancilla qubits through training parameterized rotation gates and controlled rotation gates. This method avoids the use of high-level controlled gates and the high circuit resource cost. We also extend the model to encode time-dependent potential operators by utilizing a time oracle inside the QNN circuit. Compared with the QROM-based framework, our proposed QNN-based phase kickback framework maintains accuracy in wavefunction flow simulation while significantly reducing circuit resources, enabling the implementation of flow models on quantum computers more efficiently.
    \item We combine flow matching with wavefunction flow simulation. The conventional flow model employs the velocity field learned from flow matching to generate the final distribution. In our work, we learn the potential using flow matching, and its gradient defines the conservative velocity field. Then we incorporate this velocity field into our wavefunction flow simulation. The final distribution generated by the wavefunction flow simulation successfully reconstructs the target distribution and matches the final distribution generated by the conventional flow model. This demonstrates the feasibility of employing wavefunction flow simulation for implementing flow models on quantum computers.
\end{itemize}
The remainder of this paper is organized as follows. Section~\ref{Background} provides a review of the flow model and flow matching, wavefunction flow, Hamiltonian simulation, quantum data encoding via QROM, and QNN. Section~\ref{Methodology} introduces the efficient wavefunction flow simulation through the QROM-based phase kickback framework and our proposed QNN-based phase kickback framework. Section~\ref{Results} presents the experimental results and the discussion. Section~\ref{Conclusion} provides a brief conclusion.

\section{Background}\label{Background}
\subsection{Flow model and flow matching}
The flow model is defined by a time-dependent $d$-dimensional velocity field $v(x,t)$, which generates the flow of samples through the ODE 
\begin{equation}
    \frac{d}{dt}x_t=v(x, t)\label{ODE},\; 0\le t \le 1.
\end{equation}
Given samples from the $d$-dimensional initial distribution $x_0\sim p_0$, new samples $x_1$ can be generated that follow the target $d$-dimensional distribution $p_1$ by solving Equation~\eqref{ODE}. 

Flow matching \citep{lipman2022flow} aims to learn a parameterized velocity field $v_{\theta}(x,t)$ that approximates the exact velocity field $v(x, t)$, which generates a probability path from an initial distribution $p_0$ to a target distribution $p_1$. This is done in two steps. First, a $d$-dimensional probability path $(p_t)_{0\le t\le 1}$ connecting $p_0$ and $p_1$ is defined. For samples $x_0\sim p_0$ and $x_1\sim p_1$, the flow trajectory can be viewed as the interpolation over $t\in[0,1]$ \citep{lipman2022flow, liu2022flow}
\begin{equation}
    x_t=(1-t)x_0+tx_1\label{inte}.
\end{equation}
The conditional probability path $p_t(x|x_0,x_1)$ can be defined conditioned on a sample pair $(x_0, x_1)$, resulting in the probability path that follows the expression \citep{lipman2022flow, feng2025guidance}
\begin{equation}
    p_t(x)=\int\int p_t(x|x_0,x_1)p_0(x_0)p_1(x_1)dx_0dx_1\label{probability_path},
\end{equation}
which describes the transport of probability from $x_0$ to $x_1$.

In the second step, a parameterized velocity field $v_{\theta}(x,t)$ that can generate $p_t$ is learned through neural networks. It has been demonstrated that there exists a target velocity field $v_p(x,t)$ that is capable of generating the probability path $p_t$ \citep{lipman2022flow, feng2025guidance}. This produces the flow matching objective
\begin{equation}
\mathcal{L}(\theta)_{\text{FM}}
=
\mathbb{E}_{t,x_0\sim p_0,x_1\sim p_1}
\left[
\left\|
v_\theta(x,t)-v_p(x,t)
\right\|^2
\right].
\end{equation}
However, this objective is difficult to implement, as the velocity field $v_p(x,t)$ is complicated and difficult to evaluate. Therefore, the conditional velocity field can be employed to simplify the training process \citep{lipman2022flow, lipman2024flow}. For the interpolation trajectory in Equation~\eqref{inte}, the conditional velocity field can be defined as the derivative of $x_t$ with respect to time $t$
\begin{equation}
    v(x,t|x_0,x_1)=\frac{dx_t}{dt}=x_1-x_0.
\end{equation}
Then, the conditional flow matching objective is defined as
\begin{equation}
\mathcal{L}(\theta)_{\text{CFM}}
=
\mathbb{E}_{t,x_0\sim p_0,x_1\sim p_1}
\left[
\left\|
v_\theta(x_t,t)-v(x,t|x_0,x_1)
\right\|^2
\right].
\end{equation}
A neural network with parameters $\theta$ is trained to learn $v_\theta(x_t,t)$ by minimizing the objective. After training, the trajectory can be generated using the Euler update rule
\begin{equation}
    x_{t+\Delta t}=x_t+v_\theta(x_t,t)\Delta t,
\end{equation}
or a higher-order Runge-Kutta method, such as RK45. \citep{dormand1980family}.
\subsection{Wavefunction flow}
This section reviews the wavefunction flow theory proposed in \citep{layden2025wavefunction}, including the connection between flow models and quantum dynamics, the construction of the continuity Hamiltonian, and the definition of the wavefunction flow.

Define $v_t(x):=v(x,t)$. For an infinitesimal
time step $dt$, by defining the position transformation $x_{t+dt}=x_t+v_t(x)dt$, and by employing the change of variable formula, one can describe the change of probability density through its Jacobian. This determines that the probability path $p_t$ in Equation~\eqref{probability_path}, which is generated by the velocity field $v_t(x)$, satisfies the \textit{continuity equation}
\begin{equation}
\frac{\partial}{\partial t} p_t
= - \nabla \cdot \left( v_t(x) \, p_t \right)\label{continuity_function}.
\end{equation}
The connection between flow models and quantum dynamics has been established by the definition of an appropriate continuity Hamiltonian. Given the probability path $p_t$ and its wavefunction $\Psi_t=\sqrt{p_t}$, by derivation
\begin{equation}
\begin{aligned}
i \frac{\partial}{\partial t}\Psi_t
&=
\frac{i}{2\sqrt{p_t}}
\frac{\partial}{\partial t}p_t =
-\frac{i}{2\Psi_t}
\nabla \cdot
\left(
v_t(x)\Psi_t^2
\right),
\end{aligned}
\end{equation}
and by applying the product rule, we obtain
\begin{equation}
\begin{aligned}
i \frac{\partial}{\partial t}\Psi_t
&=
-\frac{i}{2\Psi_t}
\left(
\Psi_t\,
\nabla \cdot
\left(
v_t(x)\Psi_t
\right)
+
\left(
v_t(x)\Psi_t
\right)
\cdot
\nabla \Psi_t
\right) \\
&=
\frac{1}{2}
\left(
(-i\nabla)\cdot
\left(
v_t(x)\Psi_t
\right)
+
v_t(x)\cdot
(-i\nabla)\Psi_t
\right).
\end{aligned}
\end{equation}
Thus, if $p_t$ follows the continuity equation~\eqref{continuity_function} under the velocity field $v_t(x)$, the wavefunction $\Psi_t$ follows the Schr\"odinger equation
\begin{equation}
    i \frac{\partial}{\partial t} \Psi_t = \hat{H}_t \Psi_t, \label{eq:sch}
\end{equation}
under time-dependent $d$-dimensional \textit{continuity Hamiltonian} $\hat{H}_t$ defined as
\begin{equation}
    \hat{H}_t = \frac{1}{2} \left(\hat{p} \cdot v_t(x)+ v_t(x) \cdot \hat{p} \right)\label{continuity_hamiltonian}.
\end{equation}
Here, $\hat{p}=-i\nabla$ denotes the $d$-dimensional momentum operator.

When the velocity field is conservative, i.e., $v_t=\nabla V_t$, where $V_t$ is a potential function, the corresponding time-dependent continuity Hamiltonian can be written in commutator form
\begin{equation}
\hat{H}_t = i [\hat{K}, V_t]  \label{eq:commutator_Hamiltonian}.
\end{equation}
Here, $[\hat{K}, V_t]=\hat{K}V_t-V_t\hat{K}$, and $\hat{K}=\frac{1}{2}\hat{p}\cdot \hat{p}=-\frac{1}{2}\nabla^2$ denotes the kinetic energy operator. In addition, $V_t$ represents the potential function.

Therefore, the continuity Hamiltonians defined in Equations~\eqref{continuity_hamiltonian} and~\eqref{eq:commutator_Hamiltonian} establish the equivalence between the continuity equation and the corresponding Schr\"odinger equation. This equivalence formulates the close connection between flow models and quantum dynamics. Given the initial distribution $p_0$ and the corresponding wavefunction $\Psi_0=\sqrt{p_0}$, define the entire evolution time as $T$. The final distribution $p_T$ generated by the flow model under the velocity field $v_t$ has the following relationship with the final wavefunction $\Psi_T$ evolved through the Schr\"odinger equation driven by the $v_t$-based continuity Hamiltonian
\begin{equation}
    p_T=\Psi_T^2.
\end{equation}
By constructing the continuity Hamiltonian, the quantum dynamics it generates is defined as a \textit{wavefunction flow}.
\subsection{Hamiltonian simulation}
To implement the evolution of the wavefunction flow on a quantum computer, the corresponding continuity Hamiltonian must be efficiently simulated. This can be achieved by Hamiltonian simulation, which implements quantum dynamics generated by the Schr\"odinger Equation~\eqref{eq:sch}. For the time-dependent continuity Hamiltonian $\hat{H}_t$, the solution of the Schr\"odinger equation \citep{childs2021theory} is
\begin{equation}
    \Psi_T=\mathcal{T} \exp\left(-i\int_0^T\hat{H}_\tau d\tau\right)\Psi_0,
\end{equation}
Here $\mathcal{T}$ denotes the time-ordered operator \citep{childs2021theory}, which preserves the Hamiltonians in the correct order according to the time evolution, since Hamiltonians at different times do not commute. This determines the quantum evolution from the initial state $\Psi_0$ to the final state $\Psi_T$. Therefore, the central objective of the Hamiltonian simulation is to simulate the time evolution operator 
\begin{equation}
    U(T,0)=\mathcal{T} \exp\left(-i\int_0^T\hat{H}_\tau d\tau\right)\label{operator}.
\end{equation}
The main methods for simulating the operator include the Trotter product formula \citep{low2023complexity,watson2025exponentially}, the linear combination of unitaries (LCU) \citep{berry2015simulating}, and Qubitization \citep{low2019hamiltonian}. Among all approaches, the Trotter product formula is the most straightforward to implement. Therefore, it is widely used in various scenarios. 

To approximate the operator $U(T,0)$ in Equation~\eqref{operator} using the Trotter product formula, the evolution time $T$ is generally divided into $r$ time steps with the interval $\Delta t=T/r$. Then $U(T,0)$ is expressed as
\[
U(T,0)=U\left(T,t_{r-1}\right)U\left(t_{r-1},t_{r-2}\right)\cdots U\left(t_{1},0\right),
\]
where $t_{n+1}-t_n=\Delta t$. Each decomposition operator $U(t_{n+1}, t_n)$ is defined as
\begin{equation}
U(t_{n+1}, t_n)=\mathcal{T} \exp\left(-i\int_{t_n}^{t_{n+1}}\hat{H}_\tau d\tau\right),
\end{equation}
and can be approximated by $W(t_n)$
\begin{equation}
    U(t_{n+1}, t_n)\approx W(t_n)=e^{-i\hat{H}_{t_n}\Delta t}\label{W_t}.
\end{equation}
Given the 4-step Trotter product formula for commutators \citep{chen2022efficient}
\begin{equation}
    e^{x A} e^{x B} e^{-x A} e^{-x B}
= e^{x^2[A,B]} + O(x^3),
\end{equation}
consider a Hamiltonian of the commutator form $H_{t_n}=i[A,B]$. By choosing $x=\sqrt{\Delta t}$, we can construct the corresponding approximation term $W(t_n)$, which generates 
\begin{equation}
\begin{aligned}
   W(t_n)=e^{-i\cdot i[A, B]\Delta t}&=e^{\Delta t [A, B]}\\
    &=e^{i\sqrt{\Delta t}A}
e^{i\sqrt{\Delta t}B}
e^{-i\sqrt{\Delta t}A}
e^{-i\sqrt{\Delta t}B}+O\left(\Delta t^{3/2}\right)\label{four_step}. 
\end{aligned}
\end{equation}
The 8-step product formula \citep{childs2013product} is a symmetric form of the 4-step Trotter product formula. By choosing $x=\sqrt{\frac{\Delta t}{2}}$, the formula is given by
    \begin{equation}
    \begin{aligned}
        &W(t_n)\\
        &= e^{i\sqrt{\frac{\Delta t}{2}}A}e^{i\sqrt{\frac{\Delta t}{2}}B}e^{-i\sqrt{\frac{\Delta t}{2}}A}e^{-i\sqrt{\frac{\Delta t}{2}}B}e^{-i\sqrt{\frac{\Delta t}{2}}A}e^{-i\sqrt{\frac{\Delta t}{2}}B}e^{i\sqrt{\frac{\Delta t}{2}}A}e^{i\sqrt{\frac{\Delta t}{2}}B}+O\left(\Delta t^2\right) \label{eq:symmetric_trotter}.
    \end{aligned}
\end{equation}
\subsection{Data encoding via QROM}
The question of how to load classical data into quantum computers is a critical problem in quantum algorithms. In Hamiltonian simulation, encoding the data items corresponding to the Hamiltonian operator into the quantum register is the first step. 

For example, consider a Hamiltonian that is a diagonal operator; all the diagonal entries must be encoded into the quantum register. Quantum read-only memory (QROM) is a prominent technique used to encode such classical data entries into the quantum register \citep{babbush2018encoding}; see also \citep[Sec. VI]{hann2021resilience}. Specifically, for an $n$-qubit quantum circuit system, there are $N=2^n$ basis states in total. QROM is able to encode $N$ classical data entries in the quantum register, and each data entry corresponds to a basis state. 

In the QROM oracle, each data item is represented as an $m$-bit binary string. For the $i$-th data entry $d_i$, the corresponding binary string is expressed as
\begin{equation}
    d_i=(d_{i,1},\cdots,d_{i,m}),
\end{equation}
where $d_{i,j}\in \{0,1\}$. Furthermore, $m$ ancilla qubits are used to encode these binary bits in the circuit. 

\begin{figure}[h]
\centering
\includegraphics[width=0.9\textwidth]{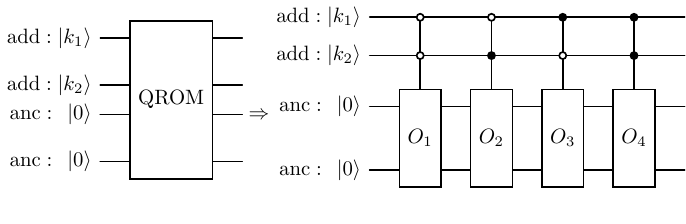}
\caption{QROM circuit for $n=2$ and $m=2$, where $\text{add}$ denotes the address qubits, and $\text{anc}$ denotes the ancilla qubits. Multi-controlled operators $O_i$ are implemented on ancilla qubits and controlled by address qubits by iterating over all $2^2=4$ address states. Black and white control nodes represent controls on $|0\rangle$ and $|1\rangle$, respectively, which select the different address states. Therefore, the controlled states of four $O_i$ operators are $|00\rangle$, $|01\rangle$, $|10\rangle$ and $|11\rangle$, respectively}\label{QROM}
\end{figure}

\noindent  For example, when $n=2$ and $m=2$, Figure~\ref{QROM} illustrates the corresponding QROM circuit. The first two qubits serve as address qubits associated with the indices of the encoded data entries. The last two qubits are ancilla qubits initialized as $|0\rangle$, which play the role of the data register. For each address $x_i\in \{0,1\}^2$, a controlled operator $O_i$ is implemented through multi-controlled $X$ gates, whose control values correspond to the address state $|x_i\rangle$. The specific implementation form of $O_i$ depends on the binary string of the $i$-th data entry $d_i$. If $d_{i,j}=1$, a multi-controlled $X$ will be implemented in the $j$-th ancilla qubit, transforming its state from $|0\rangle$ to $|1\rangle$. If $d_{i,j}=0$, no $X$ gate is applied, and the state of $j$-th ancilla qubit remains $|0\rangle$.  Hence, each controlled oracle selects a state of the address qubits and encodes the binary representation of the data entry corresponding to the same address index into the quantum circuit. 

Therefore, the operator $O_i$ performs the transformation
\begin{equation}
    |x_i\rangle |0\rangle^{\otimes m} \xrightarrow{O_i} |x_i\rangle |d_{i,1}\rangle|d_{i,2}\rangle\cdots |d_{i,m}\rangle,
\end{equation}
for all $i \in \{1, 2, \dots, N\}$. This implies that all the classical data entries are encoded into the quantum circuit via QROM.

QROM enables data encoding without preparing them in superposition \citep{crum2025stochastic}. It has been applied in quantum cryptography \citep{gidney2021factor}, chemistry \citep{lee2021even}, and some other fields to encode classical data into the quantum circuit.
\subsection{Quantum neural networks}
Quantum neural networks (QNNs) have been widely used in quantum machine learning. They combine the capabilities of both quantum computing and neural networks. QNNs are frameworks based on parameterized quantum circuits (PQCs) \citep{benedetti2019parameterized}, also known as variational quantum circuits, and optimized by classical gradient-based methods \citep{abbas2021power}. 

The workflow of a QNN is summarized as follows. First, for the $n$-qubit QNN circuit whose initial state is $|0\rangle^{\otimes n}$, the input data is encoded into the quantum circuit by preparing the corresponding quantum state $|\psi_{\text{in}}\rangle$. Then, the prepared state passes through several layers consisting of rotation gates parameterized with $\boldsymbol{\theta}=(\theta_1,\dots,\theta_m)$ and CNOT gates \citep{nielsen2010quantum}. The latter generate entanglement between qubits. The evolved state is measured by an observable, such as Pauli-$Z$. Then, the output of the QNN is defined as the expectation value of the observable \citep{zhao2021review}
\begin{equation}
    \hat{y}
=
\langle \psi_{\mathrm{out}} | M | \psi_{\mathrm{out}} \rangle,
\end{equation}
where $M$ denotes the measurement observable, and $|\psi_{\text{out}}\rangle$ represents the final quantum state generated by the QNN
\begin{equation}
    |\psi_{\mathrm{out}}\rangle
=
U(\boldsymbol{\theta})|0\rangle^{\otimes n}.
\end{equation}
The gradient of the output $\hat{y}$ with respect to $\theta_i$ can be obtained by the parameter shift rule \citep{schuld2019evaluating}. Once the gradients are obtained, we can construct a suitable loss function to measure the distance between the predicted output and the exact output. Then the learning objective is to minimize the loss function. This process can be completed outside the quantum circuit using classical gradient-based methods, such as the Adam optimizer \citep{kingma2014adam}.

Advanced architectures of QNN include quantum convolutional neural networks \citep{cong2019quantum}, quantum recurrent neural networks \citep{siemaszko2023rapid}, quantum graph neural networks \citep{verdon2019quantum}, and quantum autoencoders \citep{romero2017quantum}. Recently, the emergence of QNNs in a wide range of applications \citep{innan2024financial,mahmud2024quantum,amosy2024iteration} has helped quantum machine learning to develop rapidly in different fields. 

\section{Methodology}\label{Methodology}
In this section, first, a phase kickback framework via QROM for wavefunction flow simulation is presented (see Section~\ref{QROM-based_kickback}), followed by the proposed QNN-based phase kickback framework for more efficient simulation (see Section~\ref{QNN-based_kickback}).
\subsection{QROM-based phase kickback framework for wavefunction flow simulation}\label{QROM-based_kickback}
To implement the wavefunction flow simulation under the continuity Hamiltonian $\hat{H}_t$ defined in Equation~\eqref{eq:commutator_Hamiltonian}, both the spatial domain and the evolution time need to be discretized. In addition, we adopt a periodic spatial domain, which can be represented as a $d$-dimensional torus, to match the Fourier spatial discretization.
\subsubsection{Spatial and temporal discretization}\label{discretize}
 In this subsection, we adopt the spatial and temporal discretization methods proposed in  \citep{layden2025wavefunction}.

Since the kinetic operator is discretized through the quantum Fourier transform (QFT) \citep{nielsen2010quantum}, to match the periodic characteristic of the Fourier basis, and to avoid the spatial boundary discontinuity, the spatial domain is chosen as a torus. Consider a $d$-dimensional torus $\mathbb{T}^d$, where each spatial dimension is defined on the domain $[-L, L]$. Let $N$ be a positive integer power of 2, the domain $[-L,L]$ is first discretized into an $N$-point grid
\begin{equation}
    \mathbb{X}_N=\frac{2L}{N}\{-N/2, -N/2+1,\dots, N/2-1\}\subset \mathbb{T}^1\label{grids}.
\end{equation}
Thus, the $d$-dimensional grid $\mathbb{X}_N^d$ consists of $N^d$ spatial grid points, and each grid point $x=(x_1,\dots, x_d)\in \mathbb{X}_N^d$ corresponds to a basis state by assigning an integer index to each coordinate $x_i$
\begin{equation}
    k_i=\frac{(x_i+L)N}{2L}\in \{0,1,\dots,N-1\}.
\end{equation}
Here $k_i$ corresponds to the one-hot vector $|k_i\rangle\in \mathbb{C}^N$, i.e., $|k_i\rangle=(0,\dots,0,1,0,\cdots,0)^T$, where the $k_i$-th component is 1 and other components are 0. Therefore, the basis state corresponding to $x$ is given by
\begin{equation}
    |x\rangle
=
|k_1\rangle \otimes |k_2\rangle \otimes \cdots \otimes |k_d\rangle\label{x},
\end{equation}
where $\otimes$ denotes the tensor product of the vectors.

For the continuity Hamiltonian associated with the wavefunction flow, which is defined in Equation~\eqref{eq:commutator_Hamiltonian}, a \textit{discrete continuity Hamiltonian} is defined for simulation
\begin{equation}
    H_t=i[K,D_{V_t}].
\end{equation} 
Here, $D_{V_t}$ denotes the $N^d\times N^d$ diagonal matrix containing discretized values of the potential function $V_t$ in Equation~\eqref{eq:commutator_Hamiltonian}
\begin{equation}
    D_{V_t} = \sum_{x \in \mathbb{X}_N^d} V_t(x)\, |x\rangle\langle x|\label{DVT}.
\end{equation}
Each diagonal entry in $D_{V_t}$ corresponds to the value of $V_t$ at the associated spatial grid point $x\in \mathbb{X}_N^d$, which is given by $D_{V_t}|x\rangle=V_t(x)|x\rangle$.

The matrix $K$ denotes the discretization of the kinetic operator $\hat{K}$ in Equation~\eqref{eq:commutator_Hamiltonian}. For one spatial dimension, $\hat{K}=-\frac{1}{2}\frac{\partial^2}{\partial x^2}$. Since we have the second-order derivative of $e^{ikx}$
\begin{equation}
    \frac{\partial^2}{\partial x^2}e^{ikx}=-k^2e^{ikx},
\end{equation}
thus,
\begin{equation}
    \hat{K}e^{ikx}=-\frac{1}{2}\frac{\partial^2}{\partial x^2}e^{ikx}=\frac{1}{2}k^2e^{ikx}.
\end{equation}
Hence, $e^{ikx}$ is the eigenfunction of $\hat{K}$ with eigenvalue $\frac{1}{2}k^2$.  Therefore, the kinetic operator is diagonal in the Fourier basis. After spatial discretization, define
\begin{equation}
    k_j = \frac{\pi}{L} 
 \left( j - \frac{N}{2} \right), \;j=0,\dots, N-1,
\end{equation}
where $j - \frac{N}{2}$ represents the centered Fourier basis index. The diagonal representation of the discretized kinetic operator $K$ in the Fourier basis for one spatial dimension can be defined as
\begin{equation}
    D_K = \left( \frac{\pi}{L} \right)^2 
\sum_{j = 0}^{N-1} \left( j - \frac{N}{2} \right)^2 |j\rangle\langle j|\label{DK}.
\end{equation}
Therefore, $K$ can be expressed using $D_K$ via QFT across all $d$ spatial dimensions, which is defined as
\begin{equation}
    K = \frac{1}{2}\, (S F D_K F^\dagger S^\dagger)^{\oplus d}. \label{K}
\end{equation}
The operator $F$ represents the QFT transformation
\begin{equation}
    F=
\frac{1}{\sqrt{N}}
\sum_{j,\ell=0}^{N-1}
\exp\left(
\frac{i2\pi j\ell}{N}
\right)
|j\rangle\langle \ell|.
\end{equation}
This enables us to encode a state $|x\rangle$ by superposition of computational basis $|j\rangle$, with different relative phases and the same amplitude $\frac{1}{\sqrt{N}}$ \citep{ruiz2017quantum}
\begin{equation}
    F|x\rangle=\frac{1}{\sqrt{N}}\sum_{j=0}^{N-1}e^{i\frac{2\pi xj}{N}}|j\rangle.
\end{equation}
The phase shift operator $S$ is defined as
\begin{equation}
    S = \sum_{j = 0}^{N-1} (-1)^j\, |j\rangle\langle j|,
\end{equation}
which employ a phase shift value $(-1)^j$ to each basis state $|j\rangle$.
The operator $M^{\oplus d}$ denotes the Kronecker sum, which transforms the one-dimensional matrix $M$ into a $d$-dimensional operator
\begin{equation}
    M^{\oplus d} = (M\otimes I^{\otimes d-1}) + (I\otimes M\otimes I^{\otimes d-2}) +\cdots+ (I^{\otimes d-1}\otimes M).
\end{equation}
Each term involves applying $M$ to one dimension and applying the identity operator to other dimensions.

The temporal discretization divides the entire evolution time $T$ into multiple time steps with the interval $\Delta t$. For an interval $[t,t_1]$ with $0\le t< t_1\le T$, the 8-step Trotter product formula in Equation~\eqref{eq:symmetric_trotter} can be employed for the simulation of the operator $W(t)$ to approximate $U(t_1,t)$ defined in Equation~\eqref{W_t}. This yields
\begin{equation}
    \begin{aligned}
W(t) = {} & e^{i\beta D_{V_t}} e^{i\alpha K} 
e^{-i\beta D_{V_t}} e^{-i\alpha K} 
e^{-i\beta D_{V_t}} e^{-i\alpha K} 
e^{i\beta D_{V_t}} e^{i\alpha K}, 
\end{aligned}
\end{equation}
where
\begin{equation}
    \alpha = \frac{2L}{\pi N} \sqrt{\frac{\Delta t}{d}}, 
\qquad
\beta = \frac{\pi N}{4L} \sqrt{d\, \Delta t}\label{parameters}.
\end{equation}
The parameters satisfy $\alpha \beta = \frac{\Delta t}{2}$ according to Equation~\eqref{eq:symmetric_trotter}, and $N$ denotes the number of grids in each dimension after spatial discretization. 

The corresponding temporal discretization error $||U(t_1,t)-W(t)||$ satisfies
\begin{equation}
    \|U(t_1,t)-W(t)\|=\mathcal O\left(\Delta t^2\right),
\end{equation}
where $\|\cdot\|$ denotes the spectral norm \citep{yoshida2017spectral}. The spectral norm of a matrix $A$ is defined as
\begin{equation}
    \|A\|= \max_{\xi \neq 0}\frac{\|A\xi\|_2}{\|\xi\|_2},
\end{equation}
where $\xi$ is a non-zero vector that $A$ can act on.

The 4-step Trotter product formula for commutators defined in Equation~\eqref{four_step} can also be utilized for the approximation, which reduces the implementation of $W(t)$ from 8 exponential terms to 4. This realizes a more efficient implementation, which generates
\begin{equation}
    \begin{aligned}
W(t) = {} & e^{i\beta D_{V_t}} e^{i\alpha K} 
e^{-i\beta D_{V_t}} e^{-i\alpha K}, 
\end{aligned}
\label{Trotter} 
\end{equation}
where $\alpha$ and $\beta$ satisfy $\alpha \beta =\Delta t$ according to Equation~\eqref{four_step}. Thus, we can employ
\begin{equation}
    \alpha = \frac{2L}{\pi N} \sqrt{\frac{\Delta t}{d}}, 
\qquad
\beta = \frac{\pi N}{2L} \sqrt{d\, \Delta t}.
\end{equation}
The corresponding temporal discretization error $||U(t_1,t)-W(t)||$ satisfies
\begin{equation}
    ||U(t_1,t)-W(t)||=\mathcal O\left(\Delta t^{3/2}\right)\label{temporal_error}.
\end{equation}
\subsubsection{Phase kickback framework-based simulation}
\begin{figure}[h]
    \centering
    \includegraphics[width=0.6\linewidth]{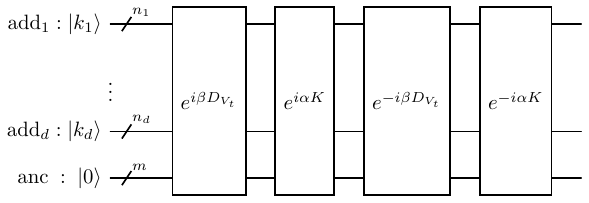}
    \caption{Quantum simulation of $W(t)$ using the 4-step Trotter product formula, where $\text{add}_j$ denotes the address qubits corresponding to the $j$-th spatial dimension, and anc denotes the ancilla qubits}
    \label{fig:Trotter}
\end{figure}

For the temporal discretization interval $[t,t_1]$, each exponential term in the corresponding approximation operator $W(t)$ defined in Equation~\eqref{Trotter} can be considered as an oracle. These oracles are implemented sequentially in the quantum circuit, as shown in Figure~\ref{fig:Trotter}. Here, $n_1,\dots, n_d$ denote the number of qubits used to represent each spatial dimension, and all of these $n=n_1+\cdots+n_d$ qubits are defined as address qubits. For the $j$-th spatial dimension, $N_j=2^{n_j}$ basis states, i.e., address states, are provided, and each address state corresponds to a discretized spatial grid point as described in Equation~\eqref{grids}. In addition, $m$ denotes the number of ancilla qubits used for data encoding.

\begin{figure}[h]
\centering
\includegraphics[width=\textwidth]{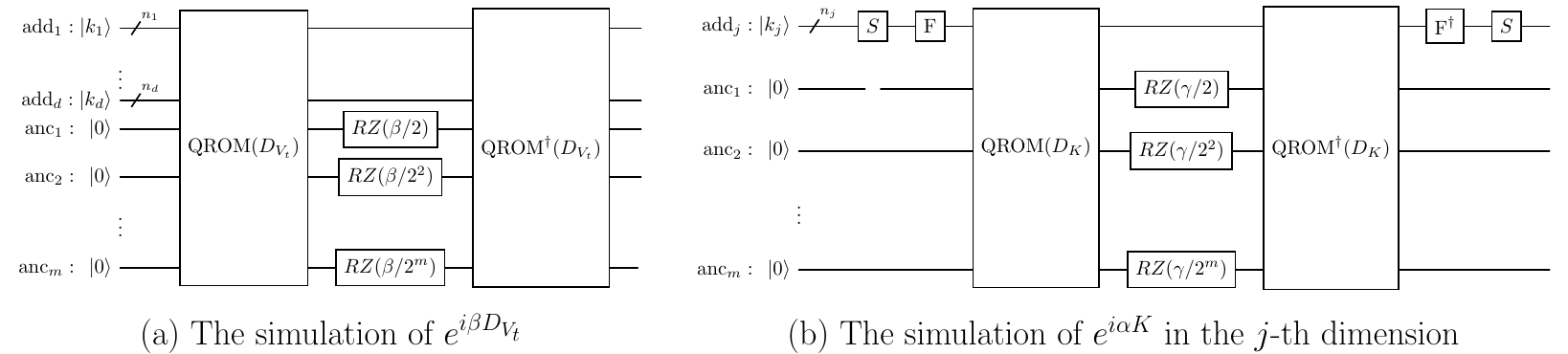}
\caption{The QROM-based phase kickback framework for the simulation of $e^{i\beta D_{V_t}}$, and the framework for the simulation of $e^{i\alpha K}$ in the $j$-th spatial dimension, where $\text{add}_j$ denotes the address qubits corresponding to the $j$-th spatial dimension, $\text{anc}_i$ denotes the $i$-th ancilla qubit, and $\gamma=\alpha/2$}\label{QROM_kickback}
\end{figure}

Each oracle is suggested to be executed using the phase kickback framework \citep{layden2025wavefunction, cleve1998quantum}. Since $D_{V_t}$ is diagonal, we incorporate QROM into the phase kickback framework to encode the diagonal entries of $D_{V_t}$ when implementing the oracle $e^{i\beta D_{V_t}}$. Figure~\ref{QROM_kickback}(a) presents the details of the phase kickback simulation of $e^{i\beta D_{V_t}}$, where the QROM$(D_{V_t})$ oracle is used to encode all diagonal entries of $D_{V_t}$ using the structure shown in Figure~\ref{QROM}.

For the $2^n\times 2^n$ diagonal matrix $D_{V_t}$ defined in Equation~\eqref{DVT}, the phase kickback framework is applied over all $n$ address qubits across all spatial dimensions. Each diagonal entry corresponds to a basis state $|k\rangle=|k_1\rangle\otimes\cdots\otimes|k_d\rangle$, which is defined in Equation~\eqref{x}. For the $l$-th min-max normalized diagonal entry $V_t^l\in [0,1]$ in $D_{V_t}$, we define its $m$-bit approximation as $0.b(t, l,1)\cdots b(t,l,m)$. Let $|s_t^l\rangle$ denote states $|b(t, l,1)\rangle\cdots |b(t,l,m)\rangle$, the QROM oracle encodes the approximation
\begin{equation}
\tilde{V}_t^l=\sum_{k=1}^{m}b(t,l,k)2^{-k},
\end{equation}
into the ancilla qubits conditioned on the corresponding address state $|l\rangle$. This realizes the state transformation
\begin{equation}
|l\rangle|0\rangle^{\otimes m}\xrightarrow{\mathrm{QROM}}|l\rangle|s_t^l\rangle.
\end{equation}
Then, an RZ$(\beta/2^{k})$ gate is applied to each ancilla qubit $k$ (see Figure~\ref{QROM_kickback}), jointly encoding the exponential $e^{i\beta\tilde{V}_t^l}$, which realizes
 \begin{equation}
 |l\rangle||s_t^l\rangle\xrightarrow{\mathrm{RZ}} e^{i\beta\tilde{V}_t^l}|l\rangle|s_t^l\rangle.
 \end{equation}
Finally, the adjoint of the QROM oracle is employed to uncompute the ancilla qubits to the initial states, which is expressed as
\begin{equation}
e^{i\beta\tilde{V}_t^l}|l\rangle|s_t^l\rangle\xrightarrow{\mathrm{QROM}^{\dagger}} e^{i\beta\tilde{V}_t^l}|l\rangle|0\rangle^{\otimes m}.\label{eq:uncompute}
\end{equation}
Thus, the phase kickback framework realizes the transformation over all basis states
\begin{equation}
\sum_k|k\rangle |0\rangle^{\otimes m}
\;\xrightarrow{\mathrm{phase\; kickback}}\;
e^{i\beta D_{V_t}}\sum_k|k\rangle|0\rangle^{\otimes m}.
\end{equation}
According to Equation~\eqref{K}, for the $N^d\times N^d$ kinetic operator $K$, to simulate the evolution operator $e^{i\alpha K}$, the phase kickback framework corresponding to $D_K$ is implemented independently to each spatial dimension. 
The implementation of $e^{i\alpha K}$ corresponding to the $j$-th spatial dimension is shown in Figure~\ref{QROM_kickback}(b), where $\gamma=\alpha/2$. Therefore, the simulation among all spatial dimensions performs
\begin{align}
    \sum_{k_1,\dots,k_d}|k_1\rangle\cdots|k_d\rangle |0\rangle^{\otimes m}
\;&\xrightarrow{\mathrm{phase\; kickback}}\;
\sum_{k_1,\dots,k_d}e^{i\alpha K}
|k_1\rangle\cdots|k_d\rangle
|0\rangle^{\otimes m}.
\end{align}
\subsection{QNN-based phase kickback framework for wavefunction flow simulation}\label{QNN-based_kickback}
\begin{figure}[h]
\centering
\includegraphics[width=\textwidth]{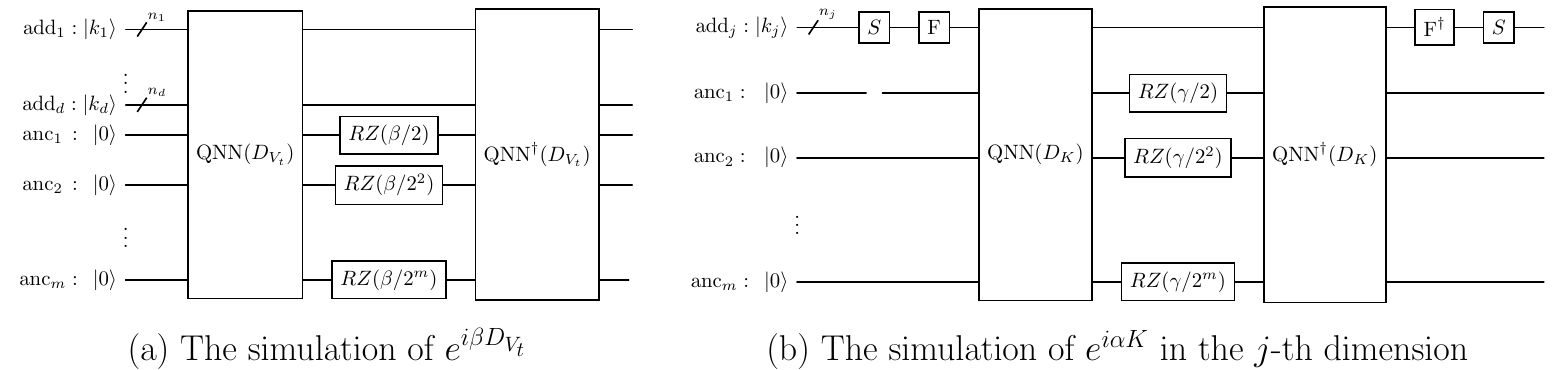}
\caption{The QNN-based phase kickback framework for the simulation of $e^{i\beta D_{V_t}}$, and the framework for the simulation of $e^{i\alpha K}$ corresponding to the $j$-th spatial dimension, where $\gamma=\alpha/2$}\label{fig:QNN_V_K}
\end{figure}

Although QROM is widely used in data encoding applications, for $n$ address qubits, it requires $\Theta(2^n)$ CNOT gates \citep{barenco1995elementary} for implementation, facing exponentially increasing computational complexity. To address this problem, we propose a QNN circuit for wavefunction flow simulation, which replaces QROM in the phase kickback framework with a trained QNN for data encoding. The proposed QNN mimics the structure of QROM by employing controlled rotation gates instead of multi-controlled $X$ gates, which reduces the number of CNOT gates to $\Theta(n)$. The structure of the QNN-based phase kickback framework to implement the potential evolution operator is illustrated in Figure~\ref{fig:QNN_V_K}(a), and the framework for implementing the kinetic evolution operator in the $j$-th spatial dimension is shown in Figure~\ref{fig:QNN_V_K}(b).
\\
We first propose a time-independent QNN circuit for the phase kickback framework of the kinetic evolution operator $e^{i\alpha K}$, which does not change over time. Then we introduce a QNN circuit for the phase kickback framework of the time-dependent potential evolution operator $e^{i\beta D_{V_t}}$ by integrating a time oracle into the time-independent QNN circuit.

\subsubsection{Time-independent QNN circuit for kinetic evolution operators}
Consider the simulation of the kinetic evolution operator $e^{i\alpha K}$ corresponding to one spatial dimension; QROM encodes the $l$-th diagonal entry $D_{K}^l$ of $D_{K}$ by loading its binary fractional
representation $0.b(l,1)\cdots b(l,m)$ into the $m$ ancilla qubits as basis states. For the $q$-th ancilla qubit, the resulting state is $|b(l,q)\rangle$ after the implementation of the QROM oracle. To replace QROM, we propose a QNN circuit, which learns the map from the address state $|l\rangle$ to the corresponding $m$ ancilla states $|b(l,1)\rangle\cdots|b(l,m)\rangle$. The QNN circuit encodes $0.b(l,1)\cdots b(l,m)$ into the $m$ ancilla qubits by loading the predicted probabilities $\hat{b}(l,1),\dots, \hat{b}(l,m)$ through minimizing the approximation error. Specifically, the trained QNN performs the map
\begin{equation}
    |l\rangle|0\rangle^{\otimes m}\xrightarrow{\mathrm{QNN}}|l\rangle|\hat{\Psi}(l)\rangle\label{QNN_map}.
\end{equation}
Here $\hat{b}(l,k)
=
P\!\left(
k\text{-th ancilla qubit}=1
\,\middle|\,
|\hat{\Psi}(l)\rangle
\right)$ denotes the predicted probability of $k\text{-th ancilla qubit}$ that is measured in state $|1\rangle$ given the state $|\hat{\Psi}(l)\rangle$ generated by the trained QNN. 

The training of the QNN can be viewed as a binary classification problem in machine learning for each ancilla qubit. Therefore, the loss function is defined using binary cross-entropy loss
\begin{equation}
L_K=
-\sum_{i=1}^{N}\sum_{h=1}^{m}
\left[
b(i,h)\log \hat b(i,h)
+
\bigl(1-b(i,h)\bigr)
\log\bigl(1-\hat b(i,h)\bigr)
\right]\label{Loss_K},
\end{equation}
where $N$ denotes the number of diagonal entries in the matrix $D_{K}$. After training, the predicted state $|\hat{\Psi}(l)\rangle$ is employed to encode the approximation of $D_{K}^l$ into the ancilla qubits by the trained QNN circuit.

\begin{figure}[h]
    \centering
    \includegraphics[width=0.8\linewidth]{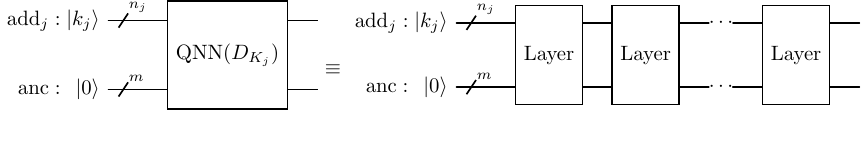}
    \caption{The QNN circuit in the phase kickback framework for the simulation of $e^{i\alpha K}$ corresponding to the $j$-th spatial dimension}
    \label{fig:QNN_K}
\end{figure}

\begin{figure}[h]
    \centering
    \includegraphics[width=0.9\linewidth]{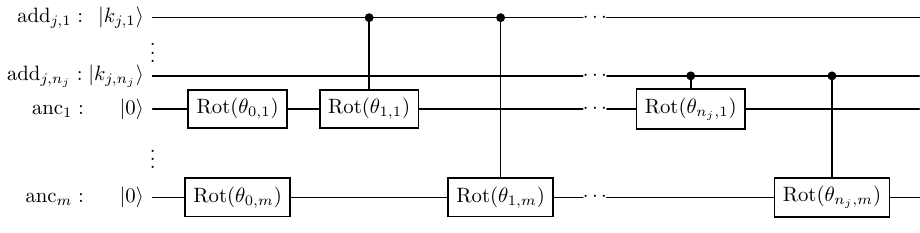}
    \caption{The QNN layer circuit in the phase kickback framework for the simulation of $e^{i\alpha {K}}$ corresponding to the $j$-th spatial dimension, where $\text{add}_{j, i}$ denotes the $i$-th address qubit corresponding to the $j$-th spatial dimension}
    \label{fig:QNN_layer_K}
\end{figure}

The time-independent QNN in each spatial dimension has repeated layers implemented in sequence, which is shown in Figure~\ref{fig:QNN_K}. Each layer contains Rot oracles and controlled-Rot oracles, which is illustrated in Figure~\ref{fig:QNN_layer_K}.

\begin{figure}[h]
    \centering
    \includegraphics[width=0.7\linewidth]{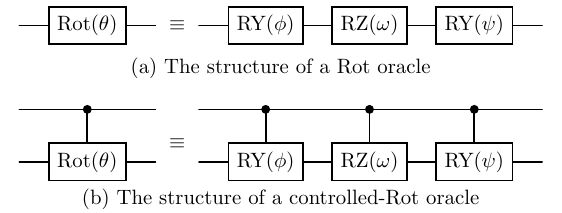}
    \caption{Circuits of Rot and controlled-Rot oracles, $\phi$, $\omega$ and $\psi$ denote three angle parameters in each oracle}
    \label{fig:Rot}
\end{figure}

Inside each layer, $m$ Rot oracles $\text{Rot}(\theta_{0,1}),\cdots, \text{Rot}(\theta_{0,m})$ are applied to the ancilla qubits in the beginning. Then, a controlled Rot oracle is applied between each address qubit and ancilla qubit. The address qubits are controls, and ancilla qubits serve as targets. Each Rot oracle consists of three rotation gates, and each controlled-Rot gate contains three controlled rotation gates, as shown in Figure~\ref{fig:Rot}.
\subsubsection{Time-dependent QNN circuit for potential evolution operators}
QROM encodes diagonal entries of the $N^d\times N^d$ diagonal matrix $D_{V_t}$ at each time $t$, respectively, where $N^d=2^n$ denotes the number of address states provided by all $n$ address qubits of $d$ dimensions jointly. We propose a time-dependent QNN circuit that can encode the diagonal operator $D_{V_t}$ at time $t$ into the phase kickback framework by incorporating a time register oracle. For a time step $h$ and the corresponding time $t_h$, given the $l$-th diagonal entry $D_{V_{t_h}}^l$ and its binary fractional representation $0.b(t_h,l,1)\cdots b(t_h,l,m)$, the QNN learns the map from the address state $|l\rangle$ to probabilities $b(t_h,l,1),\cdots, b(t_h,l,m)$. This generates
\begin{equation}
|l\rangle|0\rangle^{\otimes m}\xrightarrow{\mathrm{QNN}}|l\rangle|\hat{\Psi}(t_h,l)\rangle.
\end{equation}
Let $\hat{b}(t_h,l,k)
=
P\!\left(
k\text{-th ancilla qubit}=1
\,\middle|\,
|\hat{\Psi}(t_h,l)\rangle
\right)$ denote the predicted probability of $k\text{-th ancilla qubit}$ measured in state $|1\rangle$ for the state $|\hat{\Psi}(t_h,l)\rangle$ generated by the trained QNN.
To learn the encoding of all diagonal entries over all time steps, the loss function for QNN training is defined as
\begin{equation}
L_P=
-\sum_{k=1}^{N_t}\sum_{i=1}^{N^d}\sum_{j=1}^{m}
\left[
b(t_k,i,j)\log \hat b(t_k,i,j)
+
\bigl(1-b(t_k,i,j)\bigr)
\log\bigl(1-\hat b(t_k,i,j)\bigr)
\right]\label{Loss_V},
\end{equation}
where \(N_t\) denotes the number of discretized time intervals. After training, the predicted state $|\hat{\Psi}(t_h,l)\rangle$ is employed to encode the approximation of $D_{V_{t_h}}^l$ into the ancilla qubits by the trained QNN circuit.

\begin{figure}[h]
    \centering
    \includegraphics[width=\linewidth]{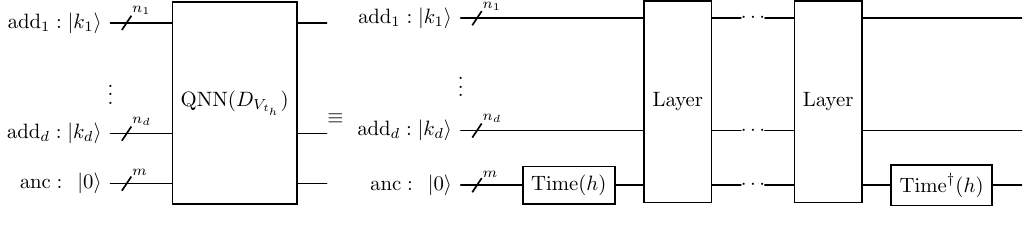}
    \caption{The time-dependent QNN circuit in the phase kickback framework for the simulation of $e^{i\beta D_{V_{t_h}}}$}
    \label{fig:QNN_V}
\end{figure}

For the time step $h$, the time-dependent QNN circuit operates on the address qubits across all spatial dimensions together with the ancilla qubits. The repeated layers are applied in sequence between a time register oracle and its adjoint oracle, which is shown in Figure~\ref{fig:QNN_V}. 

\begin{figure}[h]
    \centering
    \includegraphics[width=\linewidth]{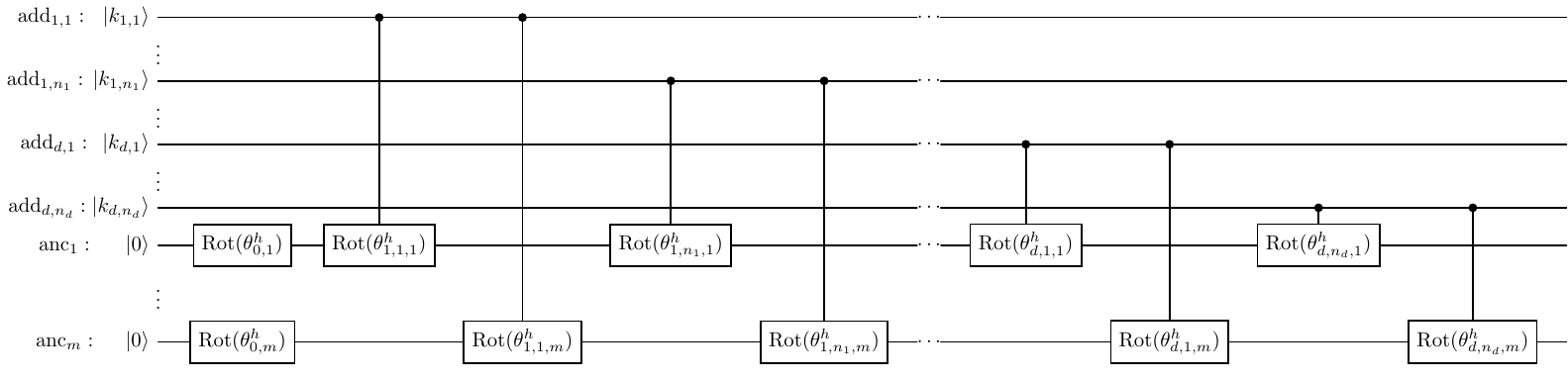}
    \caption{The QNN layer circuit in the phase kickback framework for the implementation of $e^{i\beta D_{V_{t_h}}}$}
    \label{fig:QNN_layer_V}
\end{figure}

In each layer, $m$ Rot oracles $\text{Rot}(\theta^h_{0,1}),\cdots,\text{Rot}(\theta^h_{0,m})$ operate on the $m$ ancilla qubits. A controlled-\(\mathrm{Rot}\) oracle is applied between each address qubit and the ancilla qubit. Figure~\ref{fig:QNN_layer_V} illustrates the structure of the layer.

\begin{figure}[h]
    \centering
    \includegraphics[width=\linewidth]{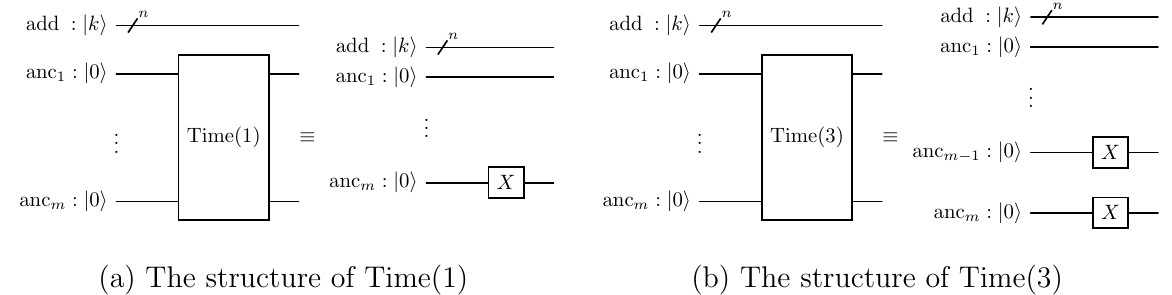}
    \caption{Examples of the Time oracles}
    \label{fig:time}
\end{figure}

The Time oracle contains $X$ gates applied to selected ancilla qubits according to the binary representation of the time step $h$. For example, when $h=0$, the states of the ancilla qubits are set as $|0\rangle^{\otimes m}$. Hence, no $X$ gate operates on the ancilla qubits. When $h=1$, the ancilla qubits are prepared in the states $|0\rangle^{\otimes (m-1)}|1\rangle$. Thus, an $X$ gate is applied to the last ancilla qubit in oracle Time(1). Examples of the Time oracle are shown in Figure~\ref{fig:time}.
\subsubsection{Resource analysis of QROM and QNN circuits}
To compare the circuit resources required for QROM and our proposed QNN simulation directly, we analyze the number of CNOT gates required in QROM and QNN circuits with the same number of address qubits and ancilla qubits.

For a QROM circuit with $n$ address qubits and $m$ ancilla qubits, whose structure is the same as shown in Figure~\ref{QROM}, the number of data entries to be encoded in the quantum register is $2^n$, corresponding to all address states. In addition, each data entry corresponds to $m$ binary bits that are encoded in ancilla qubits. Therefore, the number of multi-controlled $X$ gates in the QROM circuit is $2^nm$. 

According to the multi-controlled $X$ decomposition method without ancilla qubits \citep{barenco1995elementary}, a multi-controlled $X$ gate with $n$ control qubits consists of $3\cdot 2^{n-1}-4$ CNOT gates. Thus, the number of CNOT gates required for the QROM circuit is $(3\cdot 2^{n-1}-4)\cdot 2^n\cdot m=(3\times 2^{2n-1}-2^{n+2})m$. Therefore, for fixed $m$ and $l$, the number of CNOT gates scales as
\begin{equation}
    N_{\text{QROM}}=\Theta\left(4^n\right)\label{CNOT_QROM},
\end{equation}
resulting in exponential growth with the increase of $n$.

For our QNN circuit containing $l$ layers with the same number of address qubits and ancilla qubits, whose layer structure is shown in Figure~\ref{fig:QNN_layer_K}, $mn$ controlled-Rot oracles are required in each layer. Since each controlled-Rot oracle consists of 3 controlled rotation gates, and each controlled rotation gate requires 2 CNOT gates for decomposition \citep{barenco1995elementary}, the total number of CNOT gates in each QNN layer is $6mn$. Thus, the number of CNOT gates required for the QNN circuit is $6mnl$. Therefore, for fixed $m$ and $l$, the number of CNOT gates scales as
\begin{equation}
N_{\text{QNN}}=\Theta\left(n\right)\label{CNOT_QNN}.
\end{equation}
This shows that the number of CNOT gates grows linearly as $n$ increases in the QNN circuit.

From Equations~\eqref{CNOT_QROM} and~\eqref{CNOT_QNN}, it is clear that our proposed QNN circuit requires significantly fewer circuit resources than the QROM circuit with a large value of $n$. This is also numerically evaluated in Subsection~\ref{Circuit_Resource_Comparison}.

\section{Results and discussion}\label{Results}
\subsection{Reference method}
Given the initial distribution, to evaluate the performance of our approach, the final probability distributions generated by the wavefunction simulation are compared with the reference final distribution obtained from the corresponding conventional flow model, i.e., the solution of the continuity equation defined in Equation~\eqref{continuity_function}.

Given the initial distribution $p_0(x)$ and the corresponding wavefunction $\Psi_0(x)=\sqrt{p_0(x)}$, for the evolution time $T$, the final wavefunction $\Psi_T(x)$ generated by the wavefunction flow simulation produces the final probability distribution through $p_T(x)=|\Psi_T(x)|^2$. Here, $x=(x_1,\dots, x_d)\in \mathbb{X}_N^d$ denotes a $d$-dimensional grid point.

The reference final distribution is generated through the numerical solution of the continuity equation
\begin{equation}
\frac{\partial}{\partial t} p_t(x)
+ \nabla \cdot \left[ v_t(x) \, p_t(x) \right]=0,
\end{equation}
where the velocity field is conservative and is generated by the potential function $v_t(x)=\nabla V_t(x)$.

If the dimension of $v_t$ is 1, the continuity equation becomes

\begin{equation}
\frac{\partial}{\partial t} p_t(x)+\frac{\partial}{\partial x}(v_t(x) \, p_t(x))=0\label{continuity_function_1}.
\end{equation}
If $v_t$ is multi-dimensional, the continuity equation can be written as
\begin{equation}
\frac{\partial}{\partial t} p_t(x)
+
\sum_{i=1}^{d}
\frac{\partial}{\partial x_i}
\left(
v_{t,i}(x)
p_t(x)
\right)
=0\label{continuity_function_2}.
\end{equation}
Starting from the initial distribution \(p_0(x)\), the spatial derivative terms in the continuity function are obtained using the Fourier spectral method, and the corresponding equation is then solved from \(t=0\) to \(t=T\) using the RK45 method. This generates the reference final distribution $p^{\mathrm{ref}}_T(x)$.
\subsection{Experimental setup}
\subsubsection{Simulation environment and settings}
All quantum simulation-based experiments were conducted using Python 3.12. The quantum software stack comprised Xanadu's Pennylane 0.45.1 and pennylane-lightning 0.45.0. Quantum circuits were executed using the lightning.gpu backend on an NVIDIA Tesla T4 GPU.

The RK45 solver used to generate reference solutions of the continuity equation in Equation~\eqref{continuity_function} is provided by SciPy, and both relative and absolute tolerances are 1e-12. 

The one-dimensional and two-dimensional wavefunction flow simulations were implemented in our experiments. For the one-dimensional wavefunction flow simulation circuit, we set the number of address qubits to 4, which is capable of representing $2^4=16$ discretized spatial points in total. In addition, 10 ancilla qubits were used to encode the data entries of potential and kinetic operators into the quantum circuit. We also employed 3 address qubits for each spatial dimension when implementing the two-dimensional wavefunction flow simulation, corresponding to $2^3=8$ discretized spatial grid points per dimension. The number of ancilla qubits is 12.

In our experiments, the uniform distribution and a periodic exponential distribution given by
\begin{equation}
    p(x)=e^{-K\sin^2(\pi x/L)}
\end{equation}
were chosen as the initial distribution $p_0(x)$ for the wavefunction flow simulation and reference solutions. In the periodic exponential distribution, $K$ is a constant, and we set $K=1$ in our experiments. Both distributions are normalized so that their total probability is 1, and the corresponding initial density is represented as $\rho_0(x)$.

\subsubsection{Evaluation metrics}
Let $x^{(i)}$ denote the $i$-th discretized spatial point in the wavefunction flow simulation, and let $x^{(j)}_{\text{ref}}$ denote the $j$-th discretized spatial point in the reference method. Let $p_T(x^{(i)})$ represent the final probability at $x^{(i)}$ generated by the wavefunction flow simulation, and let $p_T^{\text{ref}}(x^{(j)}_{\text{ref}})$ denote the reference final probability at $x^{(j)}_{\text{ref}}$. Denote
\[
\rho_T(x^{(i)})=p_T(x^{(i)})/(\Delta x)^d
\]
as the corresponding wavefunction flow simulation-generated final probability density, and denote
\[
\rho_T^{\text{ref}}(x^{(j)}_{\text{ref}})=p^{\text{ref}}_T(x^{(j)}_{\text{ref}})/(\Delta x_{\text{ref}})^d
\]
as the corresponding reference final probability density. Here, $\Delta x$ is the spatial discretization interval in each dimension for wavefunction flow simulation, and $\Delta x_{\text{ref}}$ denotes the spatial discretization interval employed in the reference method. Let $N$ denote the number of discretized spatial grid points in the wavefunction flow simulation. To implement the evaluation, the reference final probability density is linearly interpolated onto the spatial grids of the wavefunction flow simulation. Let the interpolated reference final probability density at $x^{(i)}$ be represented by $\rho_T^{\text{ref}}(x^{(i)})$.

To evaluate the accuracy of the wavefunction flow simulation, we employ four error metrics: the root mean square error (RMSE), relative maximum error (RME), KL divergence ($D_{\text{KL}}$), and Wasserstein-1 distance ($W_1$)
\begin{itemize}
    \item Root mean square error (RMSE) 
    \begin{equation}
        \text{RMSE}=\sqrt{\frac{1}{N}\sum_{i=1}^N\left(\rho_T(x^{(i)})-\rho_T^{\text{ref}}(x^{(i)})\right)^2}.
    \end{equation}
    \item Relative maximum error (RME), which measures the maximum difference between probability densities $\rho_T(x)$ and $\rho_T^{\text{ref}}(x)$, normalized by the maximum value of the reference final probability density $\rho_T^{\text{ref}}(x)$
    \begin{equation}
        \text{RME}=\frac{\max_i\left|\rho_T(x^{(i)})-\rho_T^{\text{ref}}(x^{(i)})\right|}{\max_i \rho_T^{\text{ref}}(x^{(i)})}.
    \end{equation}
    \item Kullback-Leibler (KL) divergence, which measures the distance between the normalized final density $\tilde\rho_T(x)$ and the normalized reference final probability density $\tilde\rho_T^{\text{ref}}(x)$
    \begin{equation}
        D_{\text{KL}}=\sum_{i=1}^N \tilde\rho_T^{\mathrm{ref}}(x^{(i)})\ln \frac{\tilde\rho_T^{\mathrm{ref}}(x^{(i)})}{\tilde\rho_T(x^{(i)})}.
    \end{equation}
    where
    \[
    \tilde\rho_T(x^{(i)})=\frac{\rho_T(x^{(i)})}{\sum_{j=1}^N \rho_T(x^{(j)})},\quad \tilde\rho_T^{\text{ref}}(x^{(i)})=\frac{\rho_T^{\text{ref}}(x^{(i)})}{\sum_{j=1}^N \rho_T^{\text{ref}}(x^{(j)})}
    \]
    \item Wasserstein-1 distance, which measures the minimum cost of transporting the normalized final density $\tilde\rho_T(x)$ to $\tilde\rho_T^{\text{ref}}(x)$
    \begin{equation}
        W_1=\min_{\gamma_{ij} \ge 0}\sum_{i=1}^N\sum_{j=1}^N\gamma_{ij}|x^{(i)}-x^{(j)}|.
    \end{equation}
    Here
    \[
    \sum_{j=1}^N\gamma_{ij}=\tilde\rho_T(x^{(i)}),\; \sum_{i=1}^N\gamma_{ij}=\tilde\rho_T^{\text{ref}}(x^{(j)}),
    \]
and $\gamma_{ij}$ denotes the amount of probability density moving from $x^{(i)}$ to $x^{(j)}$.
\end{itemize}
Smaller values of these error metrics indicate higher consistency between the final probability densities generated by the wavefunction flow simulation and the reference final probability densities, and demonstrate higher accuracy of the wavefunction flow simulation.

\subsection{Training analysis of the QNN circuit}
To implement the wavefunction flow simulation using a trained QNN-based phase kickback framework, the QNN circuits shown in Figures~\ref{fig:QNN_K} and~\ref{fig:QNN_V} must be trained first. The values of parameters in QNN circuits were initialized from a normal distribution $\mathcal{N}(0,1)$. The evolution time $T$ was set to 0.02 s, which was divided into 200 time steps. We chose a one-dimensional potential function
\begin{equation}
V(x,t)=\cos (\pi x/L+0.75t)\label{train_analysis_v}.
\end{equation}
We analyzed the convergence of the training process and compared the performance of different numbers of QNN layers and different loss functions.
\subsubsection{Depth analysis}\label{depth_analysis}
We conducted an experiment for the convergence analysis of the QNN training process with different numbers of QNN layers. For the simulation of the potential evolution operator $e^{i\beta D_{V_{t}}}$, the QNN circuit shown in Figure~\ref{fig:QNN_V} was employed for each time step. To simulate the kinetic evolution operator $e^{i\alpha K}$, the QNN circuit shown in Figure~\ref{fig:QNN_K} was implemented independently to each spatial dimension. 

\begin{figure}[h]
    \centering
    \includegraphics[width=0.9\linewidth]{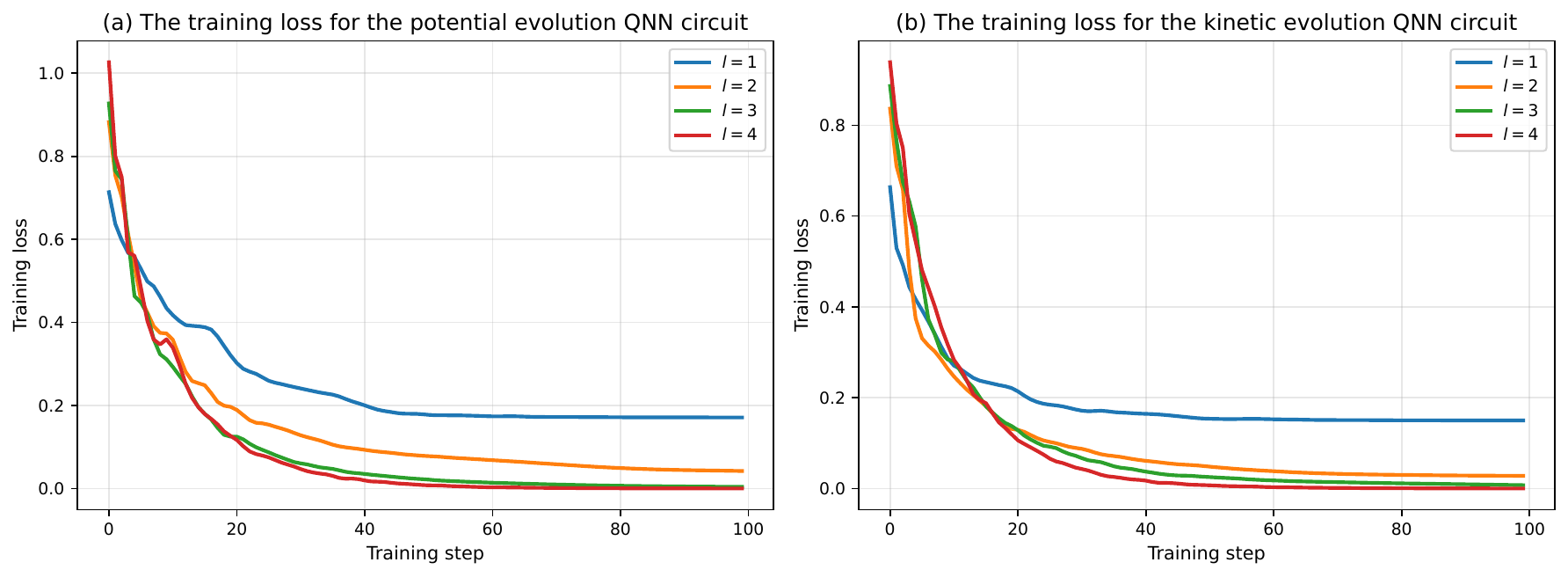}
    \caption{The training losses for the potential and kinetic evolution QNN circuits with different numbers of layers $l$}
    \label{loss_V_K}
\end{figure}

For the training of QNN circuits associated with the potential and kinetic evolution, the loss functions defined in Equations~\eqref{Loss_K} and~\eqref{Loss_V} were used, respectively. The Adam optimizer with a learning rate of 0.3 was employed in the training process. We recorded the training losses for the QNN circuits of both potential and kinetic evolution with different numbers of layers. The loss convergence curves are shown in Figure~\ref{loss_V_K}. The final loss values for potential and kinetic evolution QNN circuits with different depths are shown in Table~\ref{depth_loss}.
\begin{table}[h]
\caption{Final training loss values under different depths of QNN circuits}\label{depth_loss}%
\begin{tabular}{@{}lll@{}}
\toprule
Number of layers & QNN circuit for potential evolution  & QNN circuit for kinetic evolution \\
\midrule
1    & 0.1712   & 0.1499    \\
2    & 0.0421   & 0.0279    \\
3    & 0.0039   & 0.0073   \\
4    & 0.0001   & 0.0001   \\
\botrule
\end{tabular}
\end{table}

Figure~\ref{loss_V_K} shows that the training losses for both the QNN circuits for potential and kinetic evolution decrease as the training step increases, demonstrating the stable training process and the convergence behavior. For both QNN circuits, as the number of layers increases, the convergence of loss becomes faster, and the final training loss becomes lower. These results indicate that a deeper QNN circuit is able to improve the performance of approximating QROM in the phase kickback frameworks.

In addition, as shown in Table~\ref{depth_loss}, the final training losses of two QNN circuits for $l=1$ are 0.1712 and 0.1499, respectively. When $l$ increases to 4, the final training losses are both 0.0001. This indicates that the training loss becomes very small when the QNN depth reaches 4. In addition, as $l$ increases, the final training loss decreases rapidly, demonstrating a rapid convergence of the loss with respect to $l$.
\subsubsection{Loss function comparison}
To explore the influence of different loss functions for QNN training, we compared three loss functions in the training process
\begin{itemize}
    \item Loss function 1: Binary cross-entropy loss defined in Equations~\eqref{Loss_K} and~\eqref{Loss_V}. 
\item Loss function 2: Mean squared error between the exact values of diagonal entries and the values generated by predicted probabilities obtained from the QNN circuit
    \[
    L_K=
\sum_{i=1}^{N}\left(D_{K}^i-\sum_{h=1}^m\hat b(i,h)2^{-h}\right)^2,
\]
\[
    L_P=
\sum_{k=1}^{N_t}\sum_{i=1}^{N}\left(D_{V_{t_h}}^i-\sum_{j=1}^m\hat b(t_k,i,j)2^{-j}\right)^2.
\]
\item Loss function 3: Fidelity-based loss between the normalized vectors based on exact potential and kinetic diagonal entries and those predicted by the QNN circuit 
\[L_K
=
\frac{1}{d}
\sum_{j=1}^{d}
\left(
1-
\left|
\left\langle
\psi^{\mathrm{exact}}_{K}
\middle|
\psi^{\mathrm{pred}}_{K}
\right\rangle
\right|^2
\right),\]
    \[
    L_P=\frac{1}{N_t}
\sum_{j=1}^{N_t}
\left(
1-
\left|
\left\langle
\psi^{\mathrm{exact}}_{V_j}
\middle|
\psi^{\mathrm{pred}}_{V_j}
\right\rangle
\right|^2
\right),
    \]
where $\psi^{\mathrm{exact}}_{K}$ and $\psi^{\mathrm{exact}}_{V_j}$ denote the normalized vectors consisting of the exact diagonal entries of $D_{K}$ and $D_{V_{t_j}}$, respectively. And $\psi^{\mathrm{pred}}_{K}$ and $\psi^{\mathrm{pred}}_{V_j}$ denote the normalized vectors consisting of the predicted diagonal entries of $D_{K}$ and $D_{V_{t_j}}$ generated by the QNN circuit.
\end{itemize}

For the simulation of potential and kinetic operators, the Adam optimizer with a learning rate of 0.3 is adopted, and the number of QNN layers is 2.
\begin{figure}[h]
    \centering
    \includegraphics[width=0.9\linewidth]{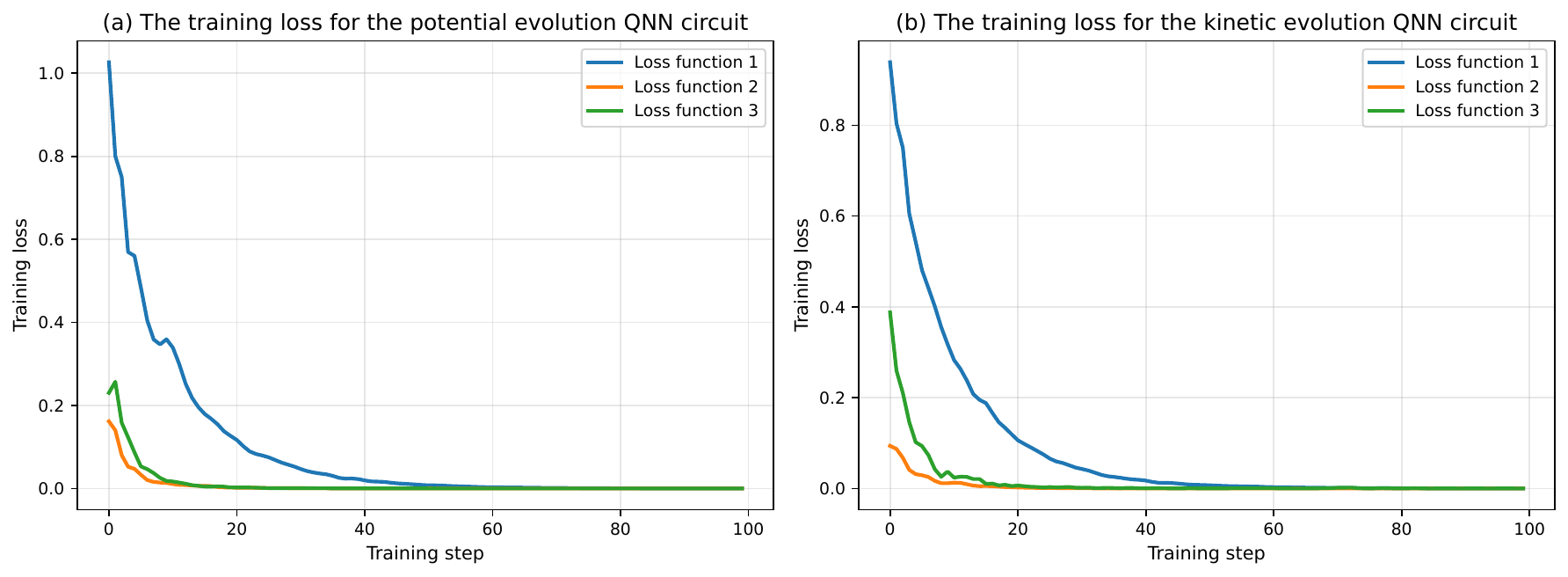}
    \caption{The training losses for the potential and kinetic evolution QNN circuits with different loss functions}
    \label{loss_V_K_function}
\end{figure}

Figure~\ref{loss_V_K_function} shows the training loss curves for both QNN circuits using three loss functions. For both QNN circuits of potential and kinetic evolution, the training losses of all three loss functions converge steadily, and their final values reach approximately 0. Meanwhile, the convergence speed of Loss function 1 is lower than those of Loss functions 2 and 3. This indicates that Loss functions 2 and 3 outperform Loss function 1 for the training process of both QNN circuits according to the training speed.

\begin{figure}[h]
    \centering
    \includegraphics[width=\linewidth]{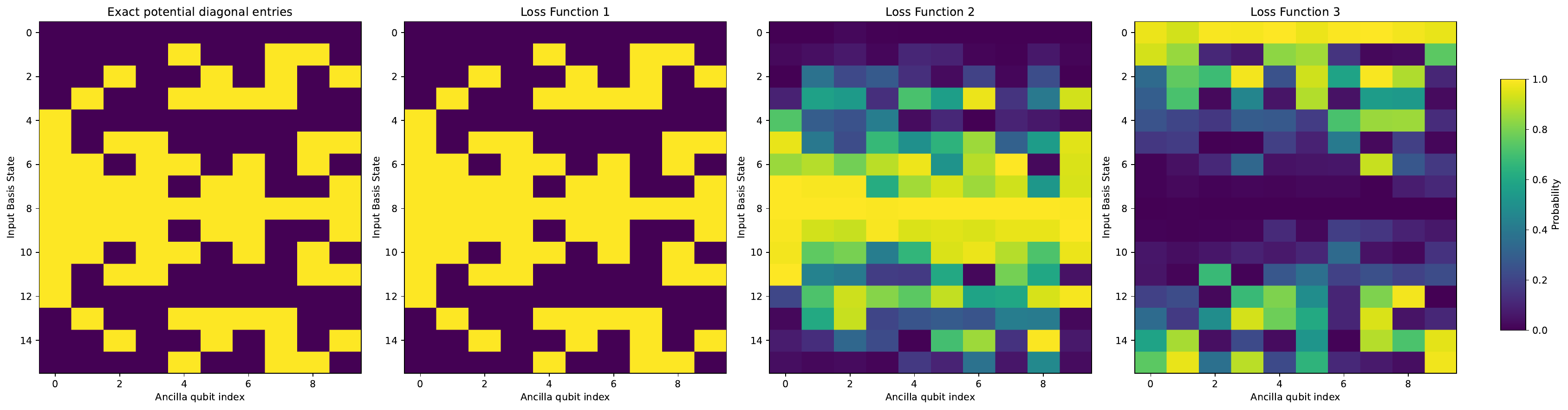}
    \caption{The comparison between the binary representations of the exact potential diagonal entries of $D_{V_{t_0}}$ and the corresponding probabilities learned by the QNN circuit under different loss functions. The $y$-axis represents the states of the input, where each state corresponds to a diagonal entry. The $x$-axis represents the ancilla qubit indices, where the $i$-th qubit index corresponds to the $i$-th binary bit of the diagonal entry}
    \label{heatmap}
\end{figure}

However, when comparing the binary representations of the exact potential diagonal entries of $D_{V_{t_0}}$ with the corresponding probabilities learned by QNN under different loss functions, different performances are exhibited. The results are shown in Figure~\ref{heatmap}. The mean distances and standard deviations across all probabilities associated with all diagonal entries of $D_{V_{t_0}}$ under different loss functions are summarized in Table~\ref{dis_bit}.
\begin{table}[h]
\caption{The mean distances and standard deviations across all probabilities under different loss functions}\label{dis_bit}%
\begin{tabular}{@{}lll@{}}
\toprule
Loss function & Mean distance &Standard deviation   \\
\midrule
Loss function 1    & 0.0001 & 0.0001      \\
Loss function 2    & 0.3726 &0.3645      \\
Loss function 3    & 0.4027 & 0.3286    \\
\botrule
\end{tabular}
\end{table}

\indent It is observed from the results that the predicted probabilities generated by the QNN circuit using Loss function 1 successfully approximate the exact binary bits of the diagonal entries, which are 0 and 1. However, the predicted probabilities generated by the QNN circuit using Loss functions 2 and 3 cannot match the corresponding binary representation, since massive probabilities stay far from 0 and 1. These indicate that, compared with the other two loss functions, the output states of the QNN circuits using the binary cross-entropy loss are precise and pure, which reduce entanglement significantly. In addition, as shown in Table~\ref{dis_bit}, the mean distance and standard deviation under Loss function 1 are both 0.0001, which are significantly lower than those obtained under Loss functions 2 and 3. These results demonstrate that the binary cross-entropy loss outperforms the other two loss functions in the training of potential and kinetic evolution QNN circuits according to the binary bit approximation.

\subsection{Comparison between QNN and QROM-based phase kickback frameworks for wavefunction flow simulation}\label{experiment 1}
To evaluate the performance of wavefunction flow simulation through our QNN-based phase kickback framework, we first compared its final probability densities with those obtained by the QROM-based phase kickback framework. Then we compared the CNOT gate cost of QROM and QNN circuits, as well as the corresponding phase kickback frameworks with different address qubits. The decomposition tools in PennyLane are used for counting the number of CNOT gates.
\subsubsection{Accuracy comparison}
We first conducted a comprehensive study of the accuracy performance of the wavefunction flow simulation using our trained QNN-based and QROM-based phase kickback frameworks under fixed potentials. To this end, given the same initial distribution, we evaluated the final probability densities generated by the two frameworks compared to the reference final densities, i.e., the results generated by the corresponding conventional flow model.

We set $L=1$, and the domain of each spatial dimension is $[-L, L]$. We chose four periodic time-dependent potentials, two of which are one-dimensional,
\[
V_1(x,t)=3\sin(\pi x/L+0.75t),
\]
\[
V_2(x,t)=3\cos(\pi x/L+0.75t).
\]
The other two potentials are two-dimensional,
\[
V_3(x,y,t)=\sin\left(\frac{\pi}{L}(x+y)+0.5t\right),
\]
\[
V_4(x,y,t)=\cos\left(\frac{\pi}{L}(x+y)+0.5t\right).
\] 
All these potentials are normalized to $[0, 1]$ using the min-max normalization.

\begin{figure}[h]
    \centering
    \includegraphics[width=\linewidth]{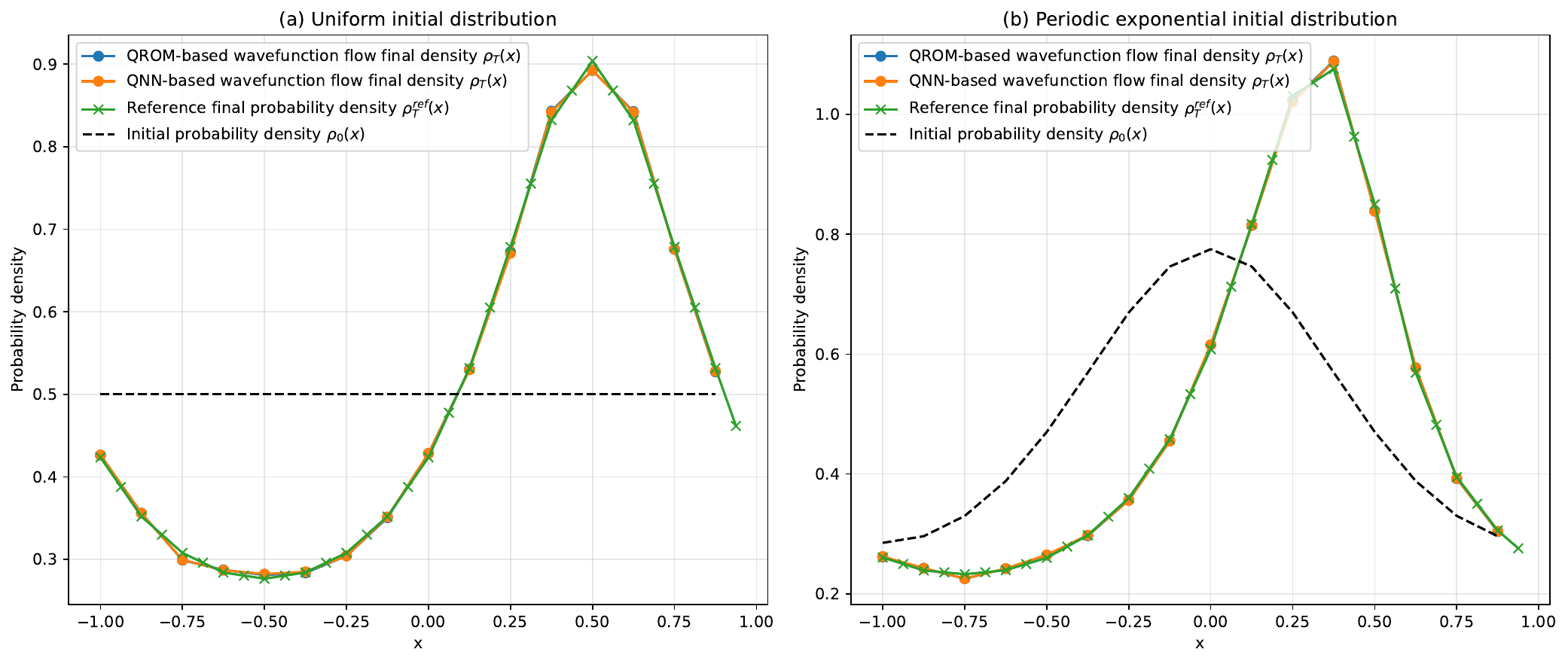}
    \caption{Comparison of final probability densities generated by wavefunction flow simulation using QNN- and QROM-based phase kickback frameworks, and reference final densities under $V_1(x,t)$}
    \label{fig:V1}
\end{figure}

\begin{figure}[h]
    \centering
    \includegraphics[width=\linewidth]{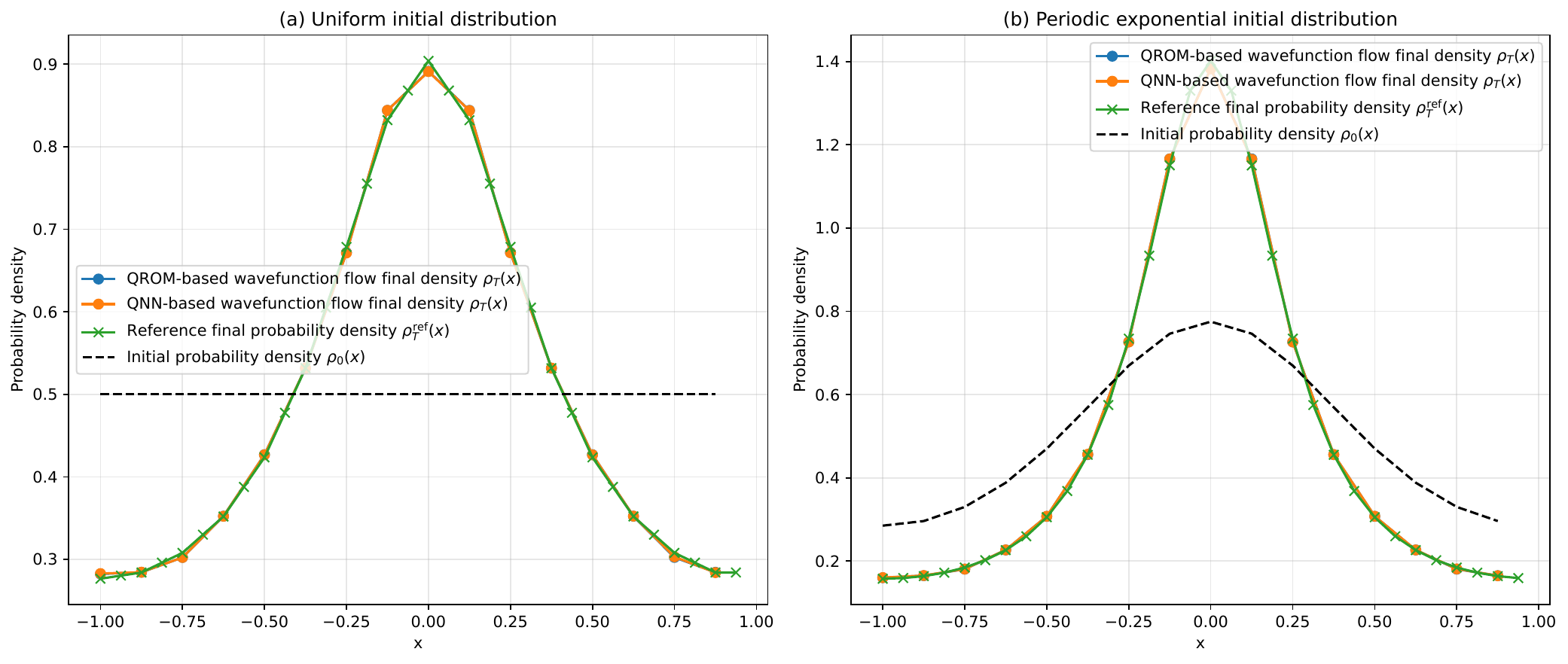}
    \caption{Comparison of final probability densities generated by wavefunction flow simulation using QNN and QROM-based phase kickback frameworks, and reference final densities under $V_2(x,t)$}
    \label{fig:V2}
\end{figure}

\begin{figure}[h]
    \centering
    \includegraphics[width=\linewidth]{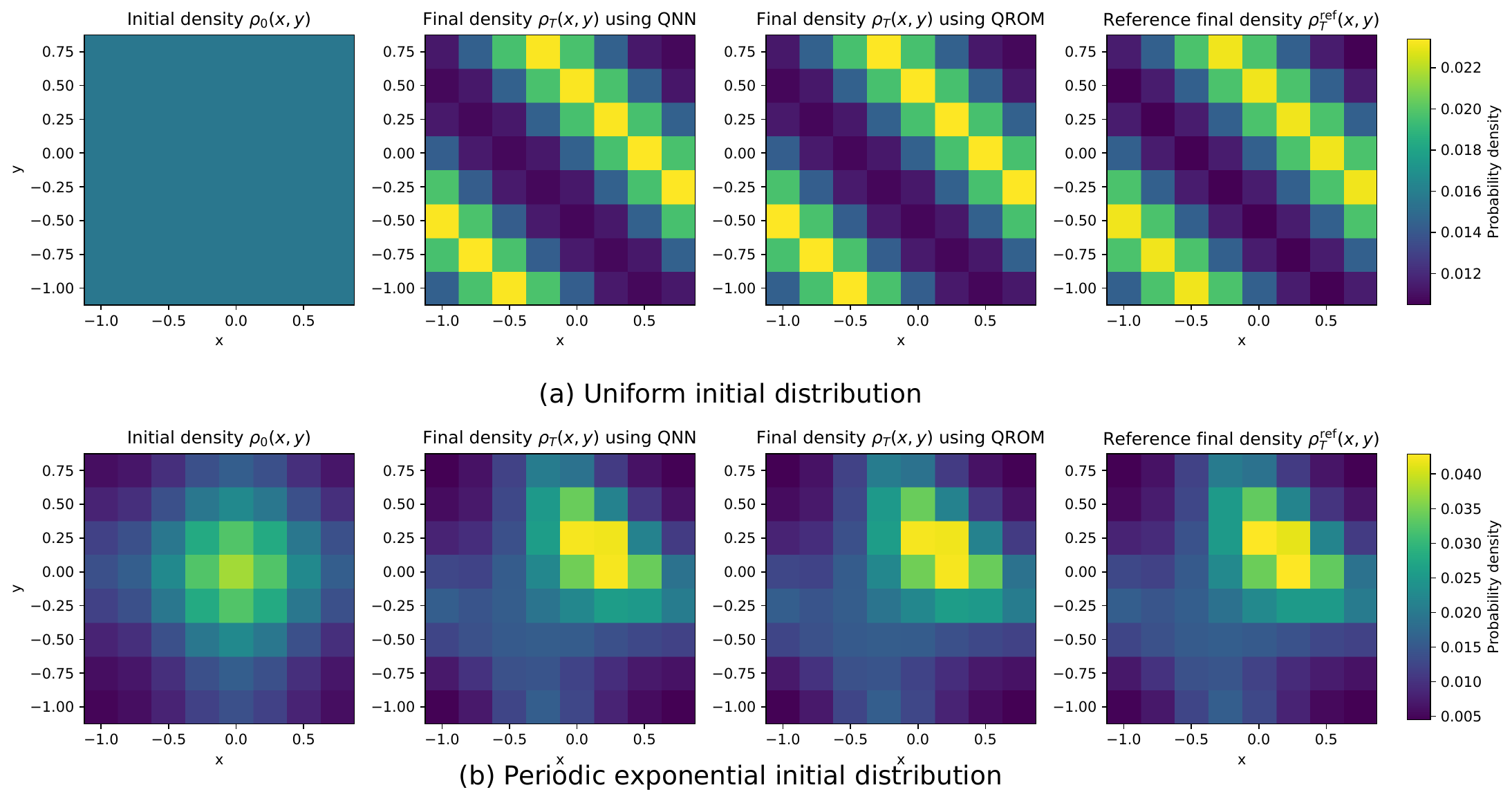}
    \caption{Comparison of final probability densities generated by wavefunction flow simulation using QNN- and QROM-based phase kickback frameworks, and reference final densities under $V_3(x,y,t)$}
    \label{fig:V3}
\end{figure}

\begin{figure}[h]
    \centering
    \includegraphics[width=\linewidth]{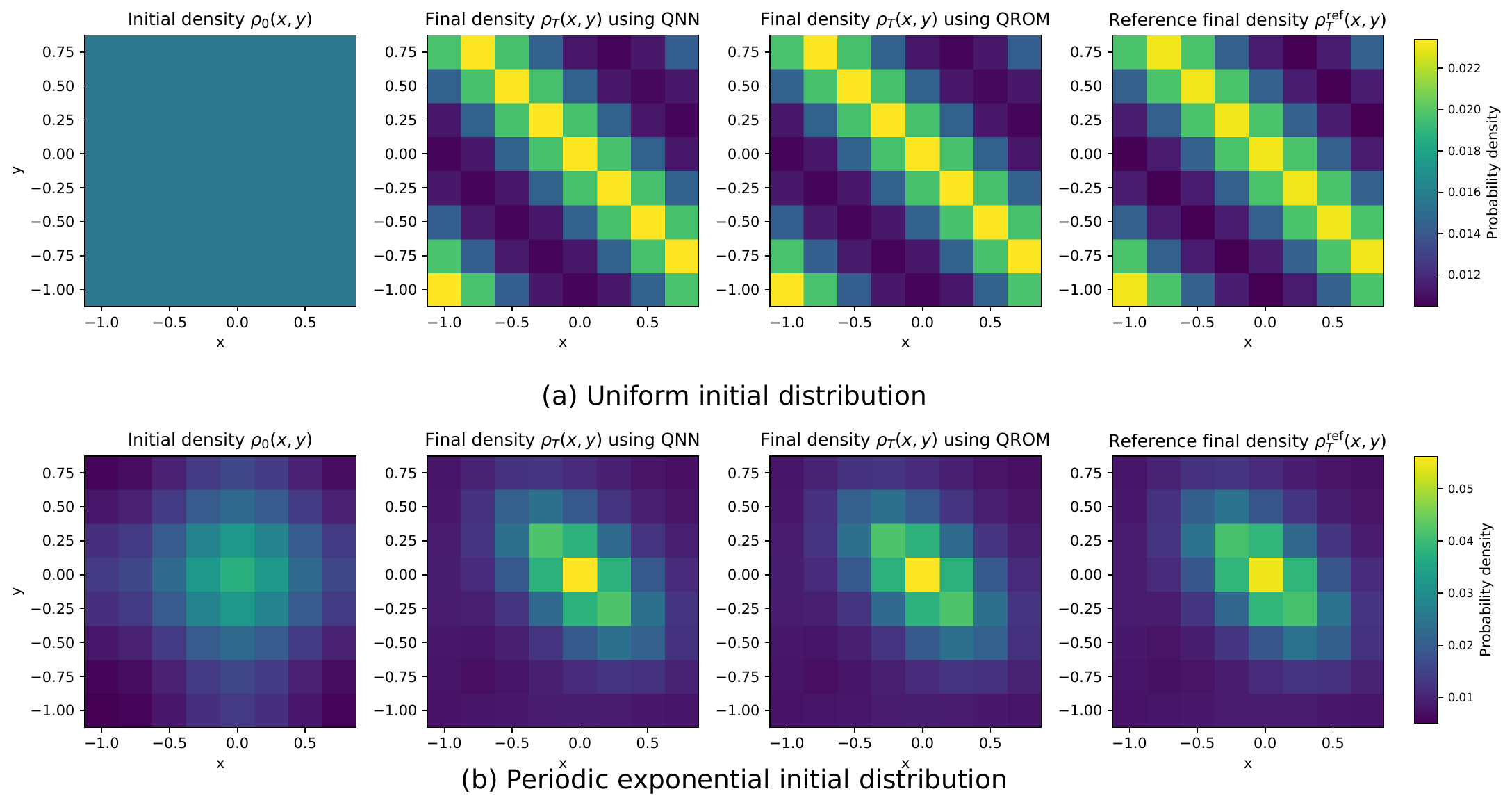}
    \caption{Comparison of final probability densities generated by wavefunction flow simulation using QNN- and QROM-based phase kickback frameworks, and reference final densities under $V_4(x,y,t)$}
    \label{fig:V4}
\end{figure}

The QROM-based phase kickback frameworks for the simulation of potential and kinetic evolution operators are shown in Figures~\ref{QROM_kickback}(a) and~\ref{QROM_kickback}(b), respectively. The evolution time $T$ was set to 0.02 s and discretized into $N_t=200$ time steps. Thus, each time interval is $\Delta t=1e-4$. We recorded the final probability densities generated by the wavefunction flow simulation using two frameworks. Then we compared them with the corresponding reference final probability densities, which were obtained by solving Equations~\eqref{continuity_function_1} and ~\eqref{continuity_function_2} using the RK45 method with the same values of $T$ and $N_t$ as those used in the wavefunction flow simulation. For the one-dimensional case, the number of discretized spatial grid points in the reference method is twice that employed in the wavefunction flow simulation. For the two-dimensional case, the number of grid points for each dimension in the reference method is the same as that in the wavefunction flow simulation. The comparison results with different potentials and initial distributions are shown in Figures~\ref{fig:V1},~\ref{fig:V2},~\ref{fig:V3}, and~\ref{fig:V4}. Tables~\ref{Error_table} and~\ref{Error_table_1} summarize the different error metrics between the final probability densities generated by the wavefunction flow simulation using QNN and QROM-based phase kickback frameworks, and the corresponding reference final probability densities.
\begin{table}[h]
\caption{Errors between final probability densities generated by wavefunction flow simulation using QNN- and QROM-based phase kickback frameworks, and reference final probability densities under the uniform initial distribution}\label{Error_table}
\begin{tabular*}{\textwidth}{@{\extracolsep\fill}lcccccccc}
\toprule%
& \multicolumn{4}{@{}c@{}}{QNN-based phase kickback} & \multicolumn{4}{@{}c@{}}{QROM-based phase kickback} \\\cmidrule{2-5}\cmidrule{6-9}%
$V$ & RMSE& RME & $D_{\text{KL}}$ & $W_1$ & RMSE& RME & $D_{\text{KL}}$ & $W_1$ \\
\midrule
$V_1$  & 5.965e-3 & 0.011&6.475e-5 & 1.119e-3 & 6.141e-3 & 0.012 & 6.677e-5&9.107e-4\\
$V_2$  & 6.424e-3 & 0.014&6.443e-5 & 9.353e-4 & 6.381e-3 & 0.014 & 6.386e-5&9.191e-4\\
$V_3$  & 1.580e-4 & 0.013&6.028e-5 & 7.397e-3 & 1.550e-4 & 0.011 & 5.306e-5&7.202e-3\\
$V_4$  & 1.572e-4 & 0.011&5.548e-5 & 6.819e-3 & 1.568e-4 & 0.011 & 5.499e-5&6.818e-3\\
\botrule
\end{tabular*}
\end{table}

\begin{table}[h]
\caption{Errors between final probability densities generated by wavefunction flow simulation using QNN- and QROM-based phase kickback frameworks, and reference final densities under the periodic exponential initial distribution}\label{Error_table_1}
\begin{tabular*}{\textwidth}{@{\extracolsep\fill}lcccccccc}
\toprule%
& \multicolumn{4}{@{}c@{}}{QNN-based phase kickback} & \multicolumn{4}{@{}c@{}}{QROM-based phase kickback} \\\cmidrule{2-5}\cmidrule{6-9}%
$V$ & RMSE& RME & $D_{\text{KL}}$ & $W_1$ & RMSE& RME & $D_{\text{KL}}$ & $W_1$ \\
\midrule
$V_1$  & 6.339e-3 & 0.012&7.010e-5 & 8.843e-4 & 6.234e-3 & 0.013 & 6.724e-5&7.669e-4\\
$V_2$  & 7.154e-3 & 0.013&5.382e-5 & 1.444e-3 & 7.852e-3& 0.013 & 6.704e-5&1.382e-3\\
$V_3$  & 2.344e-4 & 0.013&9.852e-5 & 7.870e-3 & 2.346e-4 & 0.013 & 9.826e-5&7.870e-3\\
$V_4$  & 3.465e-4 & 0.021&1.691e-4 & 1.051e-2 & 3.525e-4 & 0.022 & 1.720e-4&1.036e-2\\
\botrule
\end{tabular*}
\end{table}

We first evaluate the performance of the wavefunction flow simulation by comparing its results with those obtained from the flow model. As shown in Figures~\ref{fig:V1} and~\ref{fig:V2}, the final probability densities produced by the wavefunction flow simulation using both QNN and QROM-based phase kickback frameworks under one-dimensional potentials closely match the corresponding reference final densities under uniform and periodic exponential initial distributions. This is because their curves almost overlap and the values at each discrete spatial point nearly coincide. In addition, Figures~\ref{fig:V3} and~\ref{fig:V4} exhibit consistent densities generated by wavefunction flow simulation using two frameworks and the reference method under two-dimensional potentials. According to Tables~\ref{Error_table} and~\ref{Error_table_1}, the RMSE values corresponding to the one-dimensional potentials are on the order of $10^{-3}$, while those for the two-dimensional potentials are on the order of $10^{-4}$. These results indicate very small density errors between the wavefunction flow simulation and the reference method. The RME measures the maximum difference of all discretized spatial points between the wavefunction flow simulation and the reference method. All RME values are not greater than $0.022$, which shows that local density errors at all spatial points are well controlled.

We take small KL divergence values to indicate reliable density matching, and all the values in the two tables are below $2\times 10^{-4}$. In addition, all the values of $W_1$ in the two tables are below 0.02. These also indicate the close agreement between the densities. All these results demonstrate that the phase kickback framework-based wavefunction flow simulation generates the expected evolution of the wavefunction flow theory and matches the results of the flow model, which is obtained by solving the corresponding continuity equation.

Then we compare the accuracy of wavefunction flow using QNN- and QROM-based wavefunction flow simulation. According to Figures~\ref{fig:V1},~\ref{fig:V2},~\ref{fig:V3}, and~\ref{fig:V4}, for all four potentials, the final probability densities produced by the wavefunction flow simulation through the QNN-based phase kickback framework closely match those obtained using the QROM-based phase kickback framework. In Table~\ref{Error_table}, for $V_1$, the RMSE, RME, KL divergence, and Wasserstein-1 distance between the final probability density generated by the QNN-based phase kickback framework and the reference final probability density under uniform initial distribution are 5.965e-3, 0.011, 6.475e-5, and 1.119e-3, respectively. These values are very close to those generated by the QROM-based phase kickback framework, which are 6.141e-3, 0.012, 6.677e-5, and 9.107e-4. Similar behaviors are observed for $V_2$, $V_3$, and $V_4$ under both initial distributions. All error metrics corresponding to wavefunction flow simulation using the QNN-based phase kickback framework and the QROM-based phase kickback framework show only slight differences. These results demonstrate that the wavefunction flow simulation through the QNN-based phase kickback framework achieves accuracy comparable to that of the simulation using the QROM-based phase kickback framework.
\subsubsection{Circuit resource comparison}\label{Circuit_Resource_Comparison}
\begin{figure}[h]
\centering
\includegraphics[width=0.5\textwidth]{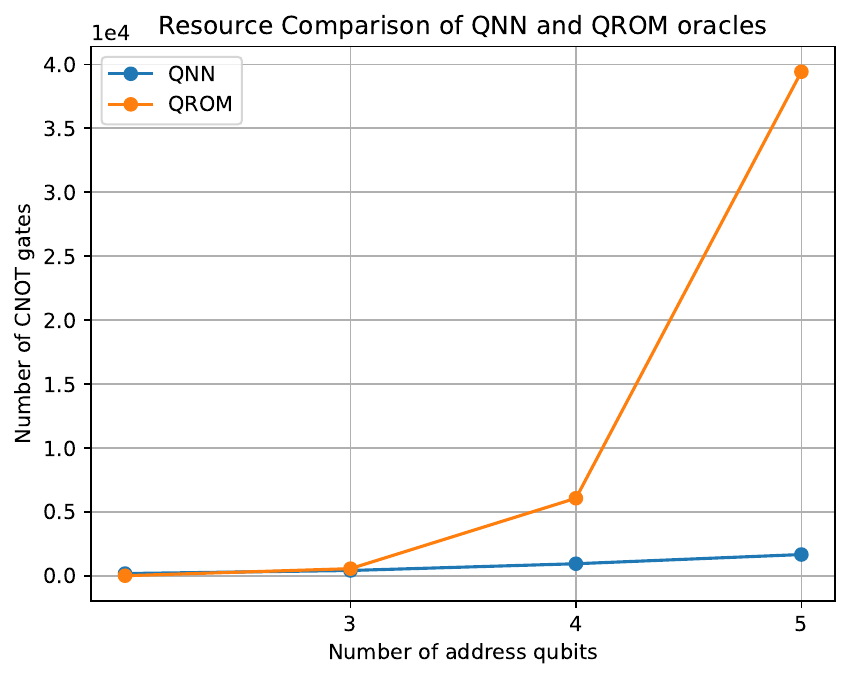}
\caption{The number of CNOT gates in QNN and QROM circuits obtained from decomposition tools in
PennyLane for the simulation of $e^{-i\beta D_{V_t}}$ corresponding to $V_1(x,t)$}\label{Resource_Comparison}.
\end{figure}

We measured the number of CNOT gates used in (i) QNN and QROM circuits, (ii) QNN-based phase kickback and QROM-based phase kickback frameworks for the simulation of $e^{-i\beta D_{V_t}}$ corresponding to the potential function $V_1(x,t)$ with different numbers of address qubits. The number of ancilla qubits was set to 6. Figure~\ref{Resource_Comparison} presents the trends of the CNOT gate counts with different numbers of address qubits. Detailed statistics for QNN and QROM circuits, as well as the corresponding phase kickback frameworks, are shown in Table~\ref{resource_table}.

\begin{table}[h]
\caption{The number of CNOT gates in QNN and QROM circuits, and the corresponding phase kickback frameworks from decomposition tools in PennyLane}
\label{resource_table}
\begin{tabular*}{\textwidth}{@{\extracolsep\fill}lcccc}
\toprule
& \multicolumn{2}{c}{QNN} & \multicolumn{2}{c}{QROM} \\
\cmidrule{2-3}\cmidrule{4-5}

$n$ & QNN circuit & Phase kickback framework & QROM circuit & Phase kickback framework \\
\midrule

2 & 192 & 384 & 24 & 48 \\

3 & 432 & 864 & 576 & 1152 \\

4 & 960 & 1920 & 6080 & 12160 \\

5 & 1680 & 3360 & 39424 & 78848 \\

\botrule
\end{tabular*}
\end{table}

As shown in Figure~\ref{Resource_Comparison} and Table~\ref{resource_table}, when $n=2$, the numbers of CNOT gates in the QNN and QROM circuits to simulate $e^{-i\beta D_{V_t}}$ are 192 and 24, respectively. As $n$ increases from 2 to 5, the number of CNOT gates in the QNN circuit increases to 1680. However, the number of CNOT gates in the QROM circuit increases exponentially to 39424, which is significantly higher than that of the QNN circuit. 

In addition, the number of CNOT gates required for the QNN-based phase kickback framework is twice that of the single QNN circuit. The same applies to the QROM-based phase kickback framework, since the phase kickback framework includes a QROM or QNN circuit together with its adjoint. 

Overall, these results demonstrate that as we introduce more address qubits to improve the precision of the wavefunction flow simulation, the QROM-based framework requires more circuit resources than the QNN-based framework. Therefore, our QNN-based framework provides a more efficient method for the implementation of flow models on quantum computers.
\subsection{Error analysis}
We investigate the errors in the phase kickback framework-based
wavefunction flow simulation. The primary sources of quantum simulation errors include the approximation errors of the values encoded in the phase kickback framework, spatial discretization errors, and temporal discretization errors.

\subsubsection{Phase kickback encoding error}
To simulate the potential evolution operator $e^{-i \beta V_t}$ using the phase kickback framework, each diagonal entry $V_j\in [0,1]$ in $D_{V_t}$ must first be encoded in the ancilla register. For a phase kickback framework with $m$ ancilla qubits, the corresponding approximation value $\tilde{V_j}$ is encoded in the quantum circuit
\begin{equation}
    \tilde{V_j}=\sum_{k=1}^{m}b_k2^{-k},
\end{equation}
where $0.b_1\cdots b_m$ denotes the binary fractional representation of $V_j$. 

\begin{figure}[h]
\centering
\includegraphics[width=0.5\textwidth]{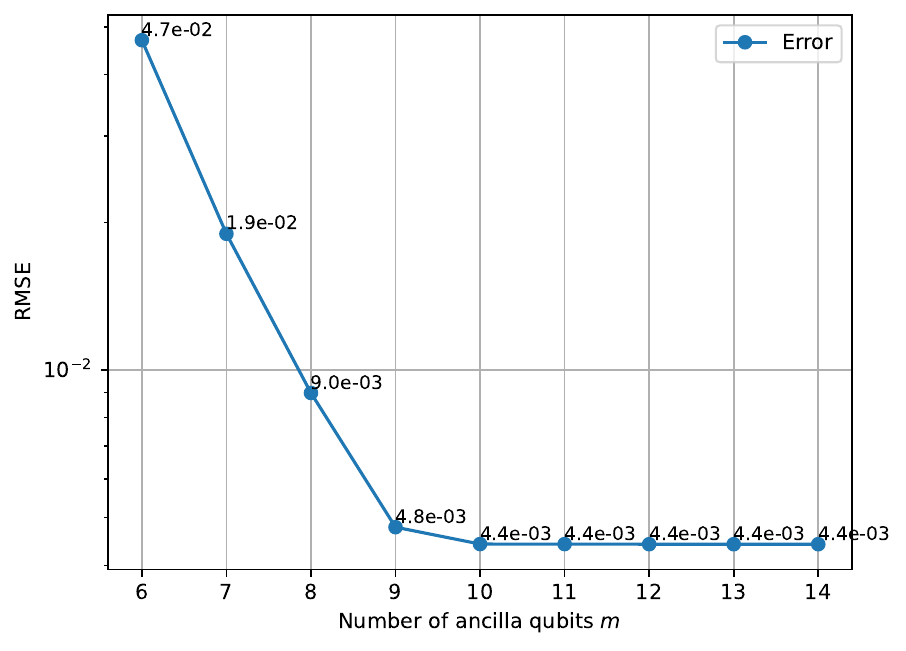}
\caption{Errors between the final probability densities generated by wavefunction flow simulation and the reference final probability densities with different numbers of ancilla qubits $m$}\label{m_error}
\end{figure}

Similarly, to simulate the kinetic evolution operator $e^{-i\alpha D_K}$, for any diagonal entry $K_j\in [0,1]$ in the operator $D_{K}$, whose binary fractional representation is $0.k_1\cdots k_m$, the approximation value $\tilde{K_j}$ is encoded in the circuit
\begin{equation}
    \tilde{K_j}=\sum_{i=1}^{m}k_i2^{-i}.
\end{equation}
When $m$ changes, the approximation errors $E_V=|\tilde{V_j}-V_j|$ and $E_K=|\tilde{K_j}-K_j|$ are expected to follow
\begin{equation}
E_V,E_K\sim 2^{-m}\label{error1}.
\end{equation}
We set the evolution time $T$ to 0.1 s, and the number of time steps to 100. The potential function was chosen to be
\begin{equation}
    V(x,t)=\sin(\pi x/L+0.75t),
    \label{V}
\end{equation}
which is normalized to $[0, 1]$ using the min-max normalization. The uniform distribution was chosen as the initial distribution. We evaluated the errors between the final densities generated by the wavefunction flow simulation and the reference final densities with different values of $m$. The results are shown in Figure~\ref{m_error}.

Increasing the number of ancilla qubits $m$ leads to an increase in the number of binary fractional representation bits of the diagonal entries. According to Equation~\eqref{error1}, the corresponding approximation error is expected to decrease. As shown in Figure~\ref{m_error}, when the number of ancilla qubits increases from 6 to 14, the final density error decreases from 4.7e-02 to a converged value of 4.4e-03. This indicates that increasing the number of ancilla qubits improves the approximation precision of the phase kickback data encoding and enhances the accuracy of the wavefunction flow simulation.

\subsubsection{Spatial discretization error}
To perform the wavefunction flow simulation corresponding to the continuous Hamiltonian $\hat{H}_t$ defined in the spatial domain $[-L, L]$, the method described in Subsection~\ref{discretize} was employed for spatial discretization. The domain of each spatial dimension was first discretized into $N$ points $\frac{2L}{N}\{-N/2, -N/2+1,\dots, N/2-1\}$. The kinetic operator corresponding to each spatial dimension was then discretized into \(N\) diagonal entries, and the potential operator was discretized into \(N^d\) diagonal entries across all $d$ dimensions. These diagonal entries are encoded in the phase kickback framework. In the wavefunction flow simulation circuit, spatial discretization depends on the number of address qubits. When the number of address qubits for each spatial dimension is $n$, the number of discretized points for each dimension is \(N=2^n\). Each discretized point corresponds to a basis state, into which the associated discretized potential and kinetic entries are encoded. 

We set the evolution time $T$ to 0.02 s, and the number of time steps $N_t$ to 200. The one-dimensional potential function was given by Equation~\eqref{V}, and the initial distribution was the uniform distribution. The number of address qubits changed from 2 to 8, corresponding to $N=[4, 8,16,32,64, 128, 256]$ discretized spatial points and discretization intervals $\Delta x=[\frac{2}{4}, \frac{2}{8}, \frac{2}{16}, \frac{2}{32}, \frac{2}{64}, \frac{2}{128}, \frac{2}{256}]$.
\begin{figure}[h]
\centering
\includegraphics[width=0.5\textwidth]{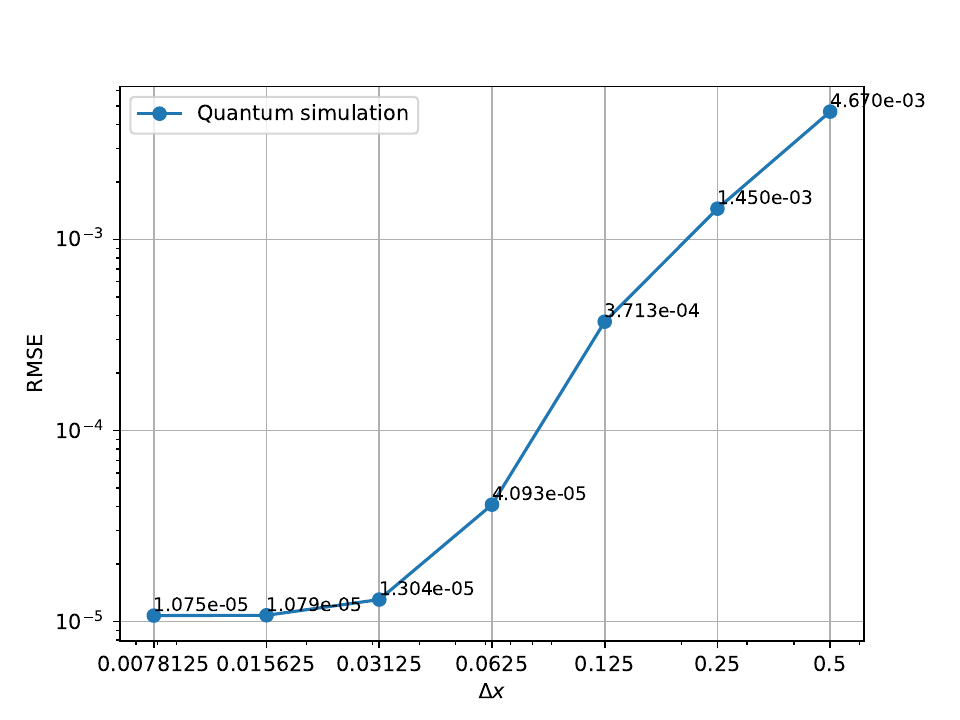}
\caption{Errors between the final probability densities generated by wavefunction flow simulation and reference final probability densities with different $\Delta x$}\label{spatial_error}
\end{figure}

Figure~\ref{spatial_error} shows the RMSE between the final probability densities generated by the wavefunction flow simulation and the reference final probability densities with different discretization intervals. When the number of address qubits $n$ increases from 2 to 5, the spatial discretization interval $\Delta x=2L/2^n$ decreases from 0.5 to 0.0625, while the RMSE decreases approximately exponentially from 4.670e-03 to 4.093e-05. This is consistent with the expected rapid decrease of the error, since we adopt Fourier spatial discretization and a smooth periodic initial distribution. As $n$ continues to increase from 5 to 8, corresponding to a decrease in \(\Delta x\) from 0.0625 to 0.0078125, the RMSE converges to a value of 1.075e-05. A further decrease of the spatial discretization interval will not lead to a lower error, since the temporal discretization error becomes dominant. These results demonstrate that finer discretized spatial intervals improve the accuracy of the wavefunction flow simulation.

\subsubsection{Temporal discretization error}
For the evolution time $T$ in the wavefunction flow simulation, the method described in Subsection~\ref{discretize} was employed for temporal discretization. The total time was divided into $N_t$ time intervals $[0,\Delta t, 2\Delta t,\dots, (N_t-1)\Delta t]$, where $\Delta t=T/N_t$ denotes the length of each interval. At each discrete time point, the diagonal operator $D_{V_t}$ varies, leading to a different quantum simulation of $e^{-i\beta D_{V_t}}$ using the Trotter product formula. The diagonal operator $D_K$ remains unchanged with time, making the simulation of $e^{-i\alpha D_{K}}$ identical at different time steps. 

We set the evolution time $T$ to 0.1 s, and the number of time steps $N_t$ varied in $N_t\in \{50, 100, 200, 250, 400, 500, 800, 1000\}$, leading to different values of the discretized temporal interval $\Delta t=T/N_t$. The initial distribution was chosen to be the uniform distribution, and the potential function was defined as
\[
V(x,t)=\left(1 + 0.5 \sin\left(0.5t\right)\right)\left(\sin(\pi x/L)\right),
\]
which is normalized to $[0, 1]$ using the min-max normalization.
\begin{figure}[h]
\centering
\includegraphics[width=0.5\textwidth]{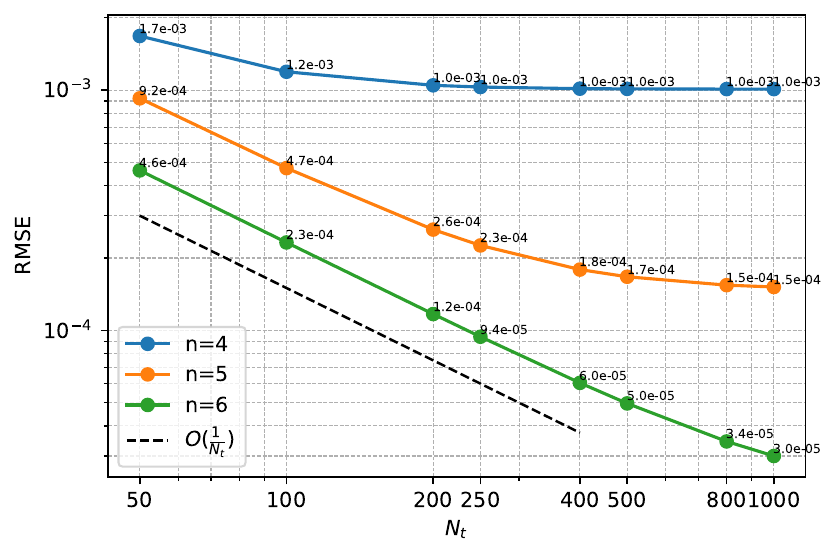}
\caption{Errors between the final probability densities generated by the wavefunction flow simulation and reference final probability densities with different $N_t$ under different numbers of address qubits}\label{dt_error}
\end{figure}\\
Figure~\ref{dt_error} exhibits the RMSE between the final probability densities generated by the wavefunction flow simulation and the reference final probability densities with different discretized temporal intervals $\Delta t$ under different discretized spatial points. We chose the number of address qubits to be 4, 5 and 6  for comparison, corresponding to the number of discretized spatial points 16, 32 and 64, respectively, leading to a decrease in $\Delta x$. As $N_t$ increases and the corresponding $\Delta t$ decreases, the final density errors under all three values of $\Delta x$ decrease. This shows that smaller temporal discretization intervals improve the precision of the wavefunction flow simulation. 

However, for $n=4$, corresponding to $\Delta x= 0.125$, the temporal discretization error converges to about 1.0e-03. For $n=5$, whose corresponding $\Delta x$ is 0.0625, the temporal discretization error converges to about 1.5e-04. This value is much lower than that of $\Delta x=0.125$. This is more significant when $n=6$, where $\Delta x$ decreases further to 0.03125, the temporal discretization error decreases linearly with increasing $N_t$ and does not show convergence until $N_t=1000$. This is consistent with a straight line scaling as $O\left(\frac{1}{N_t}\right)$. These results indicate that smaller $\Delta x$ influence the temporal discretization errors less significantly. However, we observe that the decreasing trend does not match the expected temporal discretization error with a scaling $O\left(\frac{1}{N_t^{3/2}}\right)$ as analyzed in Equation~\eqref{temporal_error}, this is because the influence of a combination of other sources of error.

\subsection{Evaluation of wavefunction flow simulation combined with flow matching}
In Subsection~\ref{experiment 1}, the potentials employed in the wavefunction flow simulation are fixed. We now consider the potential $V_t$ learned through the flow matching process.

\begin{figure}[h]
\centering
\includegraphics[width=0.9\textwidth]{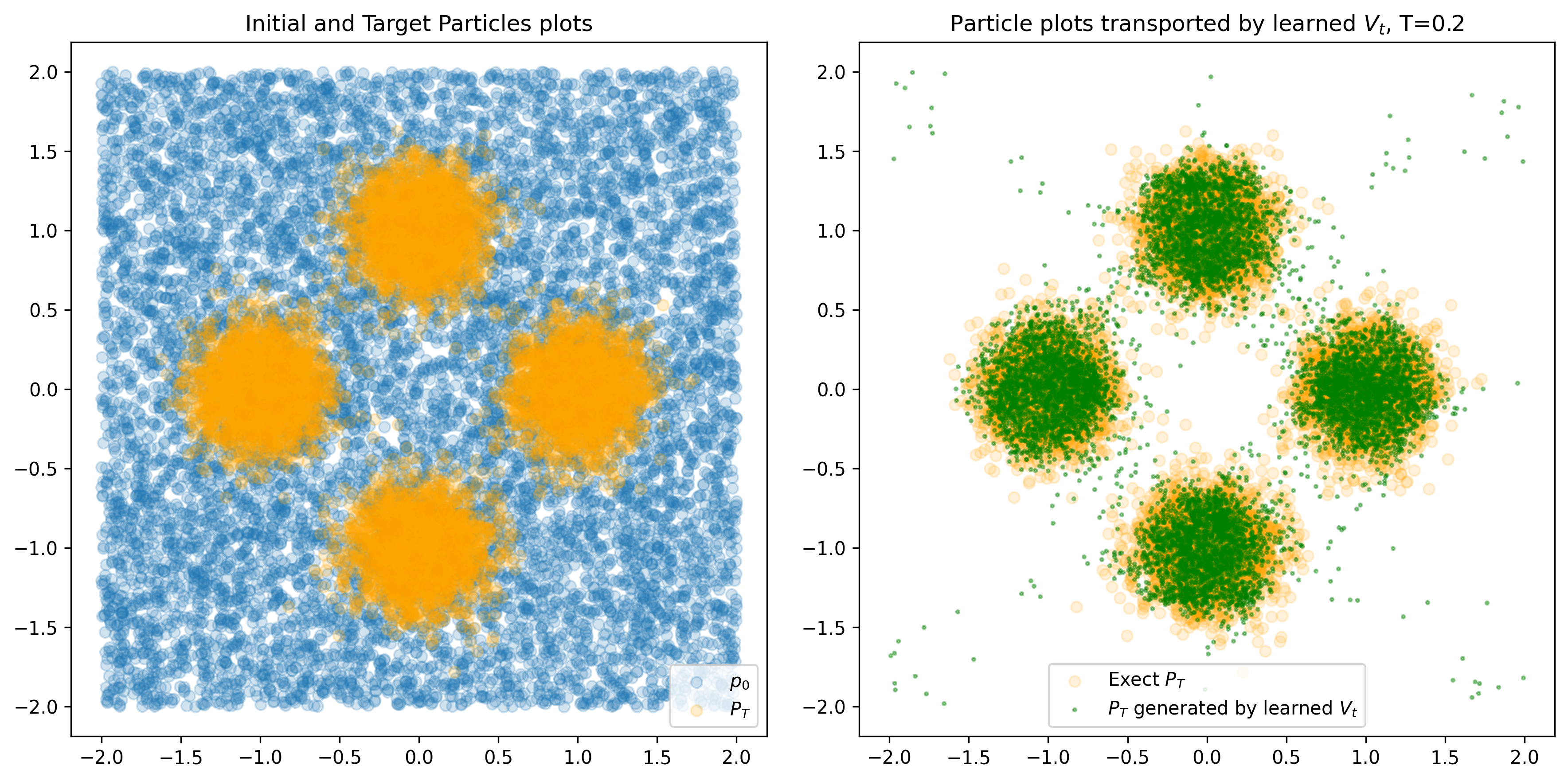}
\caption{The left figure shows the initial and the target particle plots. The right figure shows the final particle plot obtained by the flow model, in which the particles are transported under the velocity field generated by the learned potential $V_t$ through flow matching, together with the exact target particle plot}\label{Initial_target}
\end{figure}

Given the initial and target distributions, the flow matching learns a velocity field that governs the transport of particles from the initial distribution to the target distribution. We adopted a two-dimensional distribution transport experiment. The initial distribution $p_0$ is uniform in the domain $[-2,2]\times [-2,2]$, and the target distribution $p_T$ consists of four Gaussian balls centered at (1.0, 0), (-1.0, 0), (0, 1.0) and (0, -1.0), each with standard deviation 0.2. The left part of Figure~\ref{Initial_target} shows the initial and target particle plots.

Instead of directly learning the velocity field, we used neural networks to learn the potential $V_t$ using flow matching.  The gradient of $V_t$ defines the velocity field that transports particles from $p_0$ to $p_T$. The total evolution time $T$ was set to 2 s, and the number of time steps for particle transport was 2000. Then $V_t$ was learned using a fully connected neural network with layer widths of 5-128-64-32-16-1, and the Tanh activation function was employed in each hidden layer. To satisfy the periodic requirement of the spatial domain, each spatial variable was encoded using sine and cosine functions. Then they were combined with the time variable $t$, resulting in a 5-dimensional input. The velocity field generated by the learned $V_t$ was then employed to guide particle transportation starting from $p_0$ in the flow model. The right part of Figure~\ref{Initial_target} presents the final particle plot transported by the velocity field produced by the learned $V_t$, together with the target particle plot.

As shown in Figure~\ref{Initial_target}, the flow model successfully transports particles from the initial distribution to the target distribution consisting of four Gaussian balls using the learned potential through flow matching. Although there are a few scattered particles outside each Gaussian ball, they can be regarded as the error generated by flow matching learning. 

The $V_t$ learned by flow matching was then incorporated into the wavefunction flow simulation as the potential operator in the evolution of $e^{-i\beta D_{V_t}}$. The total time $T$ was set to 2 s, and the number of time steps $N_t$ was 200. The number of address qubits for each spatial dimension was set to 3, generating 8 spatial points in each dimension. The evolution was implemented following the process of Figure~\ref{fig:Trotter} with the uniform initial distribution. 

\begin{figure}[h]
\centering
\includegraphics[width=0.9\textwidth]{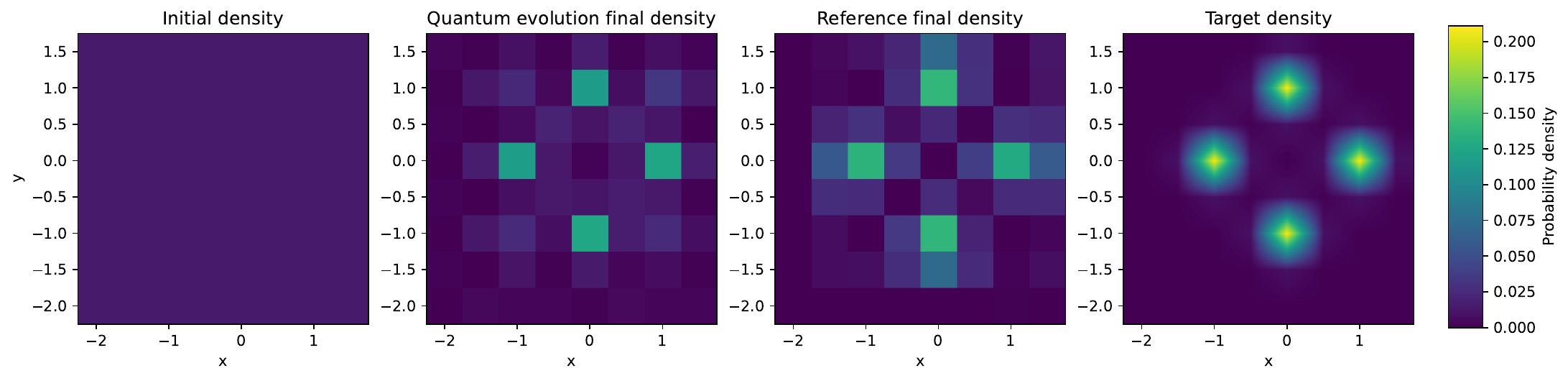}
\caption{Comparison of the final probability density obtained by wavefunction flow simulation and the target distribution. The first figure represents the initial probability density, the second figure denotes the final probability density obtained by the wavefunction flow simulation, the third figure shows the the corresponding reference final density, and the last figure shows the target probability density consisting of four Gaussian balls}\label{quantum_compare}
\end{figure}
Figure~\ref{quantum_compare} shows the initial density, the final probability density obtained by the wavefunction flow simulation, the reference final density generated by the flow model, and the target probability density. Both the final density obtained by the wavefunction flow simulation and the reference final density have four peak regions, which match the target distribution consisting of four Gaussian balls. This indicates that the wavefunction flow simulation, driven by the velocity field generated by $V_t$ learned from the flow matching, successfully generates the target density, and produces results consistent with those obtained from the flow model. Although the probability density value of each peak in the final probability density is lower than those of the reference final density and the target probability density, the similarity between the probability densities demonstrates that flow matching can be successfully integrated with the wavefunction flow simulation.

\begin{table}[h]
\caption{Comparison of final probability density errors between wavefunction flow simulation and the reference method}\label{combination_compare}%
\begin{tabular}{@{}llll@{}}
\toprule
Method & RMSE& Wasserstein-1 distance &KL divergence \\
\midrule
Wavefunction flow simulation    & 2.248e-2& 8.338e-2 & 2.935e-1   \\
Reference method    & 3.291e-2 & 9.213e-2 & 2.492e+0   \\
\botrule
\end{tabular}
\end{table}

Table~\ref{combination_compare} summarizes the RMSE, Wasserstein-1 distance, and KL divergence between the final probability density generated by the wavefunction flow simulation and the target probability density, as well as those between the corresponding reference final probability density and the target probability density. As shown in Table~\ref{combination_compare}, these values for the wavefunction flow simulation are 2.248e-2, 8.338e-2, and 2.935e-1, which are close to those generated by the reference method. These results demonstrate that the wavefunction flow simulation is comparable to the reference method, i.e. the conventional flow model, in generating the target probability density driven by the velocity field learned from flow matching.

\section{Conclusion}\label{Conclusion}
We employed a QROM-based phase kickback framework for wavefunction flow simulation to verify the wavefunction flow theory, thereby realizing the implementation of flow models on quantum computers. The experimental results show that this framework generates final distributions that are highly consistent with those obtained from the corresponding conventional flow model. We also proposed a QNN circuit to replace QROM for more efficient wavefunction flow simulation. Experiments show that our QNN-based phase kickback framework maintains the accuracy of the wavefunction flow simulation compared with the QROM-based phase kickback framework while reducing the circuit resources significantly. These results demonstrate that our approach provides a more efficient framework for implementing flow models on quantum computers. Additionally, we incorporate the potential function, which is learned by flow matching, into the wavefunction flow simulation. The results reveal that the final distribution generated by the wavefunction flow simulation successfully reproduces the target distribution and matches those obtained from conventional flow models.

Despite promising results, one major constraint of our model is that for a time-dependent potential function, the values of all discretized time steps should be included in the training process of the time-dependent QNN circuit, resulting in considerable training time. Additionally, to encode multi-dimensional potential and kinetic operators in the circuit using QNN frameworks, more than eight QNN layers are required to maintain high precision. This leads to high circuit resources. Further research could explore a more efficient QNN circuit for replacing QROM and encoding data in quantum circuits. Furthermore, we aim to develop better strategies for training the QNN circuits, which could lead to a more efficient and accurate implementation of the wavefunction flow. 

\bmhead{Acknowledgements}
We would like to express sincere gratitude to Masood Aryapoor from Mälardalen University for his help in formulating the theoretical framework, insightful discussions, and guidance in preparing the manuscript.

\section*{Statements and Declarations}

\begin{itemize}
\item \textbf{Funding} This research was partially funded by the initiative AI@MDU.
\item \textbf{Competing interests} The authors declare no competing interests.
\item \textbf{Ethics approval and consent to participate} Not applicable.
\item \textbf{Consent for publication} Not applicable.
\item \textbf{Data availability} No datasets were generated or analyzed during the current study.
\item \textbf{Materials availability} Not applicable.
\item \textbf{Code availability} The code used in this study will be made available upon reasonable request. 
\item \textbf{Author contribution} All authors conceived the research idea. Rui Che conducted the experiments and wrote the main manuscript text. Ludvig af Klinteberg provided guidance and feedback during the research process and manuscript preparation. All authors reviewed the manuscript.
\end{itemize}

\bibliography{sn-bibliography}

\end{document}